\documentclass{aa}
\usepackage{multirow}

\usepackage{numprint}
\usepackage{amsmath}
\usepackage[colorlinks,linkcolor=blue,anchorcolor=blue,citecolor=blue]{hyperref}
\usepackage{graphicx,subfig} 
\usepackage{lscape}
\usepackage{longtable}
\usepackage{txfonts}
\usepackage{natbib}
\usepackage{color}
\usepackage{booktabs}

\usepackage{natbib,twoopt}
\usepackage{xcolor}

\makeatletter
\newcommandtwoopt{\citeads}[3][][]{\href{http://adsabs.harvard.edu/abs/#3}%
    {\def\hyper@linkstart##1##2{}%
     \let\hyper@linkend\@empty\citealp[#1][#2]{#3}}}
  \newcommandtwoopt{\citepads}[3][][]{\href{http://adsabs.harvard.edu/abs/#3}%
    {\def\hyper@linkstart##1##2{}%
     \let\hyper@linkend\@empty\citep[#1][#2]{#3}}}
  \newcommandtwoopt{\citetads}[3][][]{\href{http://adsabs.harvard.edu/abs/#3}%
    {\def\hyper@linkstart##1##2{}%
     \let\hyper@linkend\@empty\citet[#1][#2]{#3}}}
  \newcommandtwoopt{\citeyearads}[3][][]%
    {\href{http://adsabs.harvard.edu/abs/#3}
    {\def\hyper@linkstart##1##2{}%
     \let\hyper@linkend\@empty\citeyear[#1][#2]{#3}}}
\makeatother
\usepackage{graphicx}
\usepackage{multirow}
\usepackage{siunitx}  
\usepackage{txfonts}

\begin{document}

\title{The eROSITA Final Equatorial-Depth Survey (eFEDS): A Machine Learning Approach to Infer Galaxy Cluster Masses from eROSITA X-ray Images}
\titlerunning{Inferring eROSITA Cluster Masses with Machine Learning}

\author{Sven Krippendorf
          \inst{1,2},
          Nicolas Baron Perez
          \inst{3,4},
          Esra Bulbul
          \inst{3},
          Melih Kara
          \inst{5},
          Riccardo Seppi
          \inst{3},
          Johan Comparat
          \inst{3},
          Emmanuel Artis
          \inst{3},
          Emre Bahar
          \inst{3},
          Christian Garrel
          \inst{3},
          Vittorio Ghiardini
          \inst{3},
          Matthias Kluge
          \inst{3},
          Ang Liu
          \inst{3},
          Miriam E.~Ramos-Ceja
          \inst{3},
          Jeremy Sanders
          \inst{3},
          Xiaoyuan Zhang
          \inst{3},
          Marcus Brüggen
          \inst{4},
          Sebastian Grandis
          \inst{6},
          Jochen Weller
          \inst{1,3}
          }

   \institute{Universit\"ats-Sternwarte, LMU Munich, Scheinerstr.~1, 81679 M\"unchen, Germany
         \and
             Arnold Sommerfeld Center for Theoretical Physics, LMU Munich, Theresienstr.~37, 80333 M\"unchen, Germany
             \and
             Max-Planck-Institut f\"ur extraterrestrische Physik, Gießenbachstraße 1, 85748 Garching, Germany
             \and
             Hamburg Observatory, University of Hamburg, Gojenbergsweg 112, 21029 Hamburg, Germany
             \and
             Institute for Astroparticle Physics, Karlsruhe Institute of Technology, 76021 Karlsruhe, Germany
             \and
             Institute for Astro- and Particle Physics, University of Innsbruck, Technikerstr.~25, 6020 Innsbruck, Austria
             }
\authorrunning{Krippendorf, Baron Perez et al.}

\abstract{
\noindent We develop a neural network based pipeline to estimate masses of galaxy clusters with a known redshift directly from photon information in X-rays. Our neural networks are trained using supervised learning on simulations of {\it eROSITA} observations, focusing in this paper on the Final Equatorial Depth Survey (eFEDS). We use convolutional neural networks which are modified to include additional information of the cluster, in particular its redshift. In contrast to existing work, we utilize simulations including background and point sources to develop a tool which is usable directly on observational {\it eROSITA} data for an extended mass range from group size halos to massive clusters with masses in between $10^{13}M_\odot<M<10^{15}M_\odot.$ Using this method, we are able to provide for the first time neural network mass estimation for the observed eFEDS cluster sample from Spectrum-Roentgen-Gamma/{\it eROSITA} observations and we find consistent performance with weak lensing calibrated masses. In this measurement, we do not use weak lensing information and we only use previous cluster mass information which was used to calibrate the cluster properties in the simulations. When compared to simulated data, we observe a reduced scatter with respect to luminosity and count-rate based scaling relations. 
 We comment on the application for other upcoming {\it eROSITA} All-Sky Survey observations. 
}
\maketitle
\newpage

\section{Introduction}
Improving our understanding of the mass function of galaxy clusters enables us to improve our inference on key cosmological parameters. These parameters include $\Omega_M$, the density parameter of matter in the Universe, and $\sigma_8$ which describes the dispersion of linear density fluctuations.
The ongoing {\it eROSITA} (extended ROentgen Survey with an Imaging Telescope Array)  All-Sky Survey~\citep{2021A&A...647A...1P}  on board the Spectrum Roentgen Gamma mission \citep{Sunyaev2021} will provide us with the largest intra-cluster medium (ICM)-selected galaxy clusters to date which promises to provide tight constraints on cosmology through cluster abundance measurements ~\citep{2012arXiv1209.3114M}.
A key ingredient in this analysis is to understand the cluster masses associated with a selected underlying sample \citep{Bulbul2019}. Traditionally this is performed with weak lensing (WL) calibrated scaling relations in the context of the eROSITA cluster census or using dynamical mass measurements~\citep{2013MNRAS.429.3079M, 2014MNRAS.441.1513O,2015MNRAS.449.1897O} in situations where data allows for this approach. In the context of {\it eROSITA}, the former procedure has been demonstrated on the Final Equatorial Depth Survey (eFEDS) using the Hyper-Supreme Camera WL mass measurements ~\citep[see][]{Bahar2022, chiu}.

Cosmology analyses through cluster abundances detected in the X-ray or SZ surveys heavily rely on the availability of external WL mass measurements \citep{Mantz2015, Bocquet2019,Grandis:2018mle}. This procedure requires the knowledge of cluster masses through WL surveys and introduces bias and scatter in the final cosmology contours if survey data are not deep enough. Unaccounting for these biases and selection differences may affect the final cosmology measurements \citep{Ramos-Ceja2022}. Recently, applications of new machine learning (ML) tools and methods on large astronomy data and numerical simulations presented a promising method to reduce scatter on such cluster mass calibration using X-ray images \citep[see][]{ntampaka1, Green_2019, Yan_2020}, SZ Compton $y-$maps \citep{Cohn_2019, 2201.01305, 2209.02075}, and using optical data \citep{Ntampaka_2015, Ho_2019, Kodi_Ramanah_2020, Ho_2021, Ho_2022}.

In this work, we present a method that avoids the explicit knowledge of these WL measurements by using X-ray data and the redshift of clusters. In spirit, this is the same approach as using existing scaling relations on a new cluster sample. To calibrate, or, put differently, train our ML model we are using simulations and the accuracy of these methods is determined by the cluster model in the training data. To apply this method on new observations reliably, we are interested in training our ML model with a realistic cluster sample, i.e.~simulated clusters which represent our knowledge on clusters based on previous observations and represent the observational setting. In comparison to standard scaling relations, this ML model is more flexible as it can combine different features in a non-linear model. Furthermore, we consider models that utilize most of the information available, including the observation's energy and spatial information, rather than preprocessed features such as the luminosity (profiles) of a galaxy cluster. Given the success in other domains with similar data structures (such as in computer vision tasks on images~\cite{krizhevsky2017imagenet}), a natural candidate for such models are convolutional neural networks (CNNs). The potential of these methods for estimating galaxy cluster masses has been previously demonstrated in~\cite{ntampaka1}, where a reduced mass scatter compared to luminosity-based methods was reported. In this work, we modify these methods to address a cluster sample at a larger redshift ($0.01<z<1.3$) and mass range ($10^{13}<M/M_\odot<10^{15}$). Additionally, we account for emissions from other X-ray sources, e.g., active galactic nuclei (AGN), that are major contaminators in cluster analyses. Here, we present a method where additional filtering for such point sources is not required.

\begin{figure*}
\centering
\includegraphics[width=\textwidth]{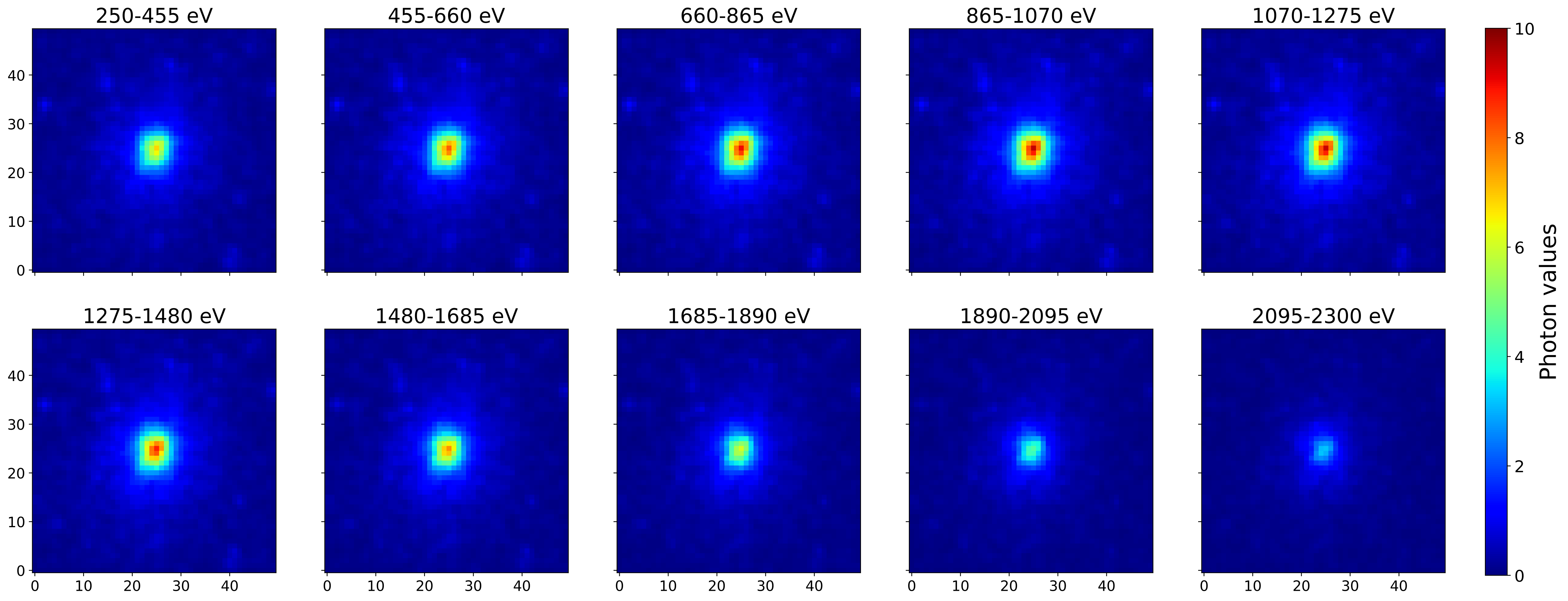}
\caption{An example X-ray image as input to our neural networks.  All ten bands of the image of the galaxy cluster (SRC\_ID: 10006566 from realization $5$) of the \textit{eFEDS} simulations. Each image has a dimension of $50\times 50.$ Due to our smoothing, the photon values are continuous. This cluster has a mass $\log{\left(M_{500}/M_\odot\right)}=15$ and is located at a redshift of 0.11.}
\label{fig:energybandimage}
\end{figure*}

Finally, our neural network (NN) method incorporates a measure of uncertainty alongside the respective mass prediction. To estimate the uncertainty, we assume that the logarithm of our cluster masses is distributed according to some underlying Gaussian distribution with an associated mean and standard deviation. Both can be inferred using the log-likelihood associated with this normal distribution (cf.~Section~\ref{sec:ml-method} for a detailed description). In addition, to account for the model uncertainty of our NN, we use a frequentist ensemble approach for our final mean and standard deviation. We train and validate our method on simulations of {\it eROSITA} galaxy clusters dedicated for eFEDS observations~\citep{comparat-simulation,Liu:2021idr, Seppi2022}.

This allows us to apply our method on the eFEDS cluster sample~\citep{liu-efeds-clusters} and provide for the first time ML mass estimates on cluster observations. When comparing the performance of our mass estimates with those obtained from WL calibrated scaling relations using count-rate measurements \citep{chiu} on the simulations we find a reduced scatter. Our results on simulations are of similar scatter as using idealised luminosity information.

The paper is organized as follows: In Section~\ref{sec:data}, we describe the respective data products used in this work. Section~\ref{sec:ml-method} describes our machine learning approach, and we discuss the results of our numerical work in Section~\ref{sec:results}. Our conclusions are presented in Section~\ref{sec:conclusions}.

Throughout this paper, our simulated observations are obtained using a flat $\Lambda$CDM cosmology close to that of the Planck collaboration~\cite{2020A&A...641A...6P} with $H_0 = 67.74$~km~s$^{-1}$~Mpc$^{-1},$ $\Omega_{\mathrm m} =0.308900$, $\Omega_{\mathrm b} =0.048206$, and $\sigma_8 = 0.8147$ as described in~\cite{comparat-simulation}. 
Our masses $M_{500c}$ refer to the mass included in the region with a mean density of 500 times the critical density.

\section{{\it eROSITA} X-ray and Simulated Observations}
\label{sec:data}

This section presents the data we have used to train and test our machine learning method. We restrict ourselves to the data corresponding to the performance verification mini-survey of {\it eROSITA}, eFEDS \citep{Brunner2022}, the data analysis pipeline~\citep{Liu:2021idr} and the corresponding eFEDS simulations (see~\cite{Comparat_2019} for the procedure on how AGNs are simulated, \cite{comparat-simulation} how galaxy clusters are painted for $M_{500c}>10^{13.7}M_{\odot}$ and~\citet{Seppi2022} for the extension to lower masses $M_{500c}>10^{13}M_{\odot}$).

We comment on the extension of our method to {\it eROSITA} All-Sky survey (eRASS) observations in our conclusions in Section~\ref{sec:conclusions}. 

\subsection{{\it eROSITA} X-ray Images}

The 140~deg$^2$ eFEDS field, designed as a performance verification survey, has a uniform depth of 2.2~ks (1.2~ks after correcting for vignetting) approximately equal to the depth of the final {\it eROSITA} All-Sky Survey \citep{Brunner2022}. In this field, a total of 542 cluster candidates were detected with an extent likelihood threshold larger than six and detection likelihood larger than five \citep[see][for details]{liu-efeds-clusters}. Of these 542 candidates, 477 galaxy groups and clusters were confirmed with the follow-up optical data with redshift measurements \citep{Klein2022}. The clusters detected in the point source catalog are excluded in this analysis due to differences in the selection criteria \citep{Bulbul2022}. 
We use the subsample of 463 optically confirmed clusters, which have WL-calibrated features between $10^{13}M_{\odot}<M_{500}<10^{15}M_{\odot}.$ 
This selection is applied as this corresponds to the mass range on which our networks are trained on, i.e.~the cluster sample from the eFEDS simulations subsequently described.

To create X-ray images, we use the {\it eROSITA} Standard Analysis Software System \citep[{{\tt eSASS}}][]{Brunner2022}, version \texttt{eSASSusers\_201009}. The calibrated event lists are corrected for good time intervals, dead times, corrupted events and frames, and bad pixels. Images are generated in ten equally spaced energy intervals of $205~{\rm eV}$ each in the soft band for the range $0.25-2.30~{\rm keV}$, 
using the {\tt eSASS} tool {\tt evtool}. Multiple energy bands are selected to maximize the information on the X-ray images, taking advantage of the superb soft sensitivity of {\it eROSITA}. We keep X-ray photons in a fixed square of 300 pixels (corresponding to $1200''$) centered on the X-ray centroid identified by~{\tt eSASS}. 

\subsection{{\it eROSITA} Simulated Images}

The mock observations used in this study have the same exposure depth and field area to match the eFEDS observations. A method developed in \citep{comparat-simulation, Seppi2022} is employed to generate the mock photons for our training, validation, and test sets. A full-sky dark matter-only simulation provides the halo sample. Based on the properties of the dark matter halos, the X-ray properties of the sources are impainted using a Gaussian process model, which has been fit using previous cluster observations. These properties are then used to generate a source list passed to the SIXTE software \citep{dauser2019sixte}, which outputs the survey mock photons. It is worth stressing that these include not only cluster photons but, in addition, also point sources.

Within the eFEDS simulation, 18 realizations of the same eFEDS field are created to have enough sources for statistical analysis \citep[see][]{liu-efeds-clusters}. All realizations together contain $\numprint{148833}$ clusters, whereas a single realization contains approximately $\numprint{8000}$ clusters. To train our neural networks on a representative sample, we restrict ourselves to the same thresholds for cluster selection used for the eFEDS catalog (\texttt{eSASS} software version \texttt{eSASSusers\_211214}). This gives a final sample of $\numprint{7947}$ clusters. We found that these selection criteria improved our ML performance compared to using more clusters utilizing the ones with smaller detection and extent likelihoods. As for the observations, we use X-ray photons in a fixed square of $300$ pixels centered around the halo cluster center, noting no performance difference between the simulated cluster center and the eSASS detected center.

\subsection{Neural network input datasets}

These respective images are in a standard data format for images in machine learning which are processed as a three-dimensional array where the first two dimensions carry the spatial information and the third dimension respectively carries ``color" information. We modify the images to make our machine-learning pipeline more efficient.

To render the input less sparse and to have fewer memory requirements, we scale the boxes of size $300\times 300 \times 10$ down to a size of $50\times 50 \times 10$, and we apply Gaussian smoothing in all three directions, including the energy direction. The respective formula can be found in Appendix~\ref{app:data}.

We do not remove background photons or identified point sources, but we clip the pixel number at $36$ to avoid instabilities in our neural network training. 
An example of such an energy-band-image (EBI) can be found in Figure~\ref{fig:energybandimage}. To resolve the ambiguity between a less luminous cluster at low redshift and a highly luminous cluster at high redshift, we also use the redshift information as input to our network. We optimized the spatial region and smoothing used for the EBI, ensuring that in almost all cases, the entire cluster is visible in the image.

It is important to stress that these input images are independent of $R_{500c}$ as such a selection would automatically include information about the cluster mass.

\begin{figure}[t]
\begin{center}
\includegraphics[width=0.45\textwidth]{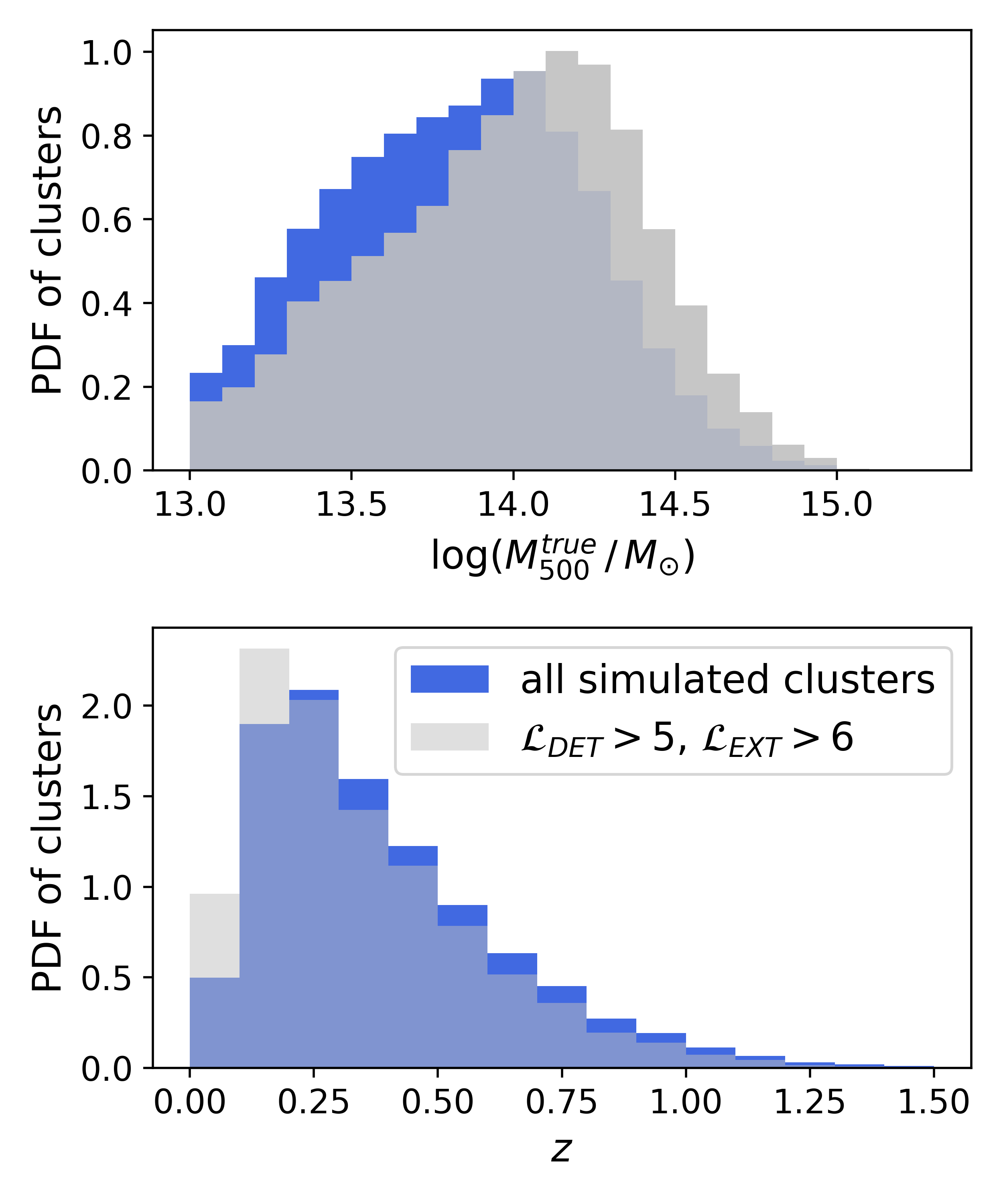}
\end{center}
\caption{The simulated cluster sample with and without the applied filters used for training and evaluation with their respective redshift and mass distribution. The total number of simulated and filtered clusters is 25031 and 7947, respectively.}
\label{fig:distributions}
\end{figure}

The respective mass and redshift distributions of clusters in eFEDS simulations and our {\tt eSASS} selected sample are shown in Figure~\ref{fig:distributions}. From all realizations of the simulations, we end up with $7947$ clusters from which we use $70\%$ for our training, $15\%$ for our validation, and $15\%$ for our test set.

\section{Machine Learning Method}
\label{sec:ml-method}

\begin{figure*}[t]
\begin{center}
\includegraphics[width=0.68\textwidth]{images/CNN_Architecture-logL.pdf}
\includegraphics[width=0.3\textwidth]{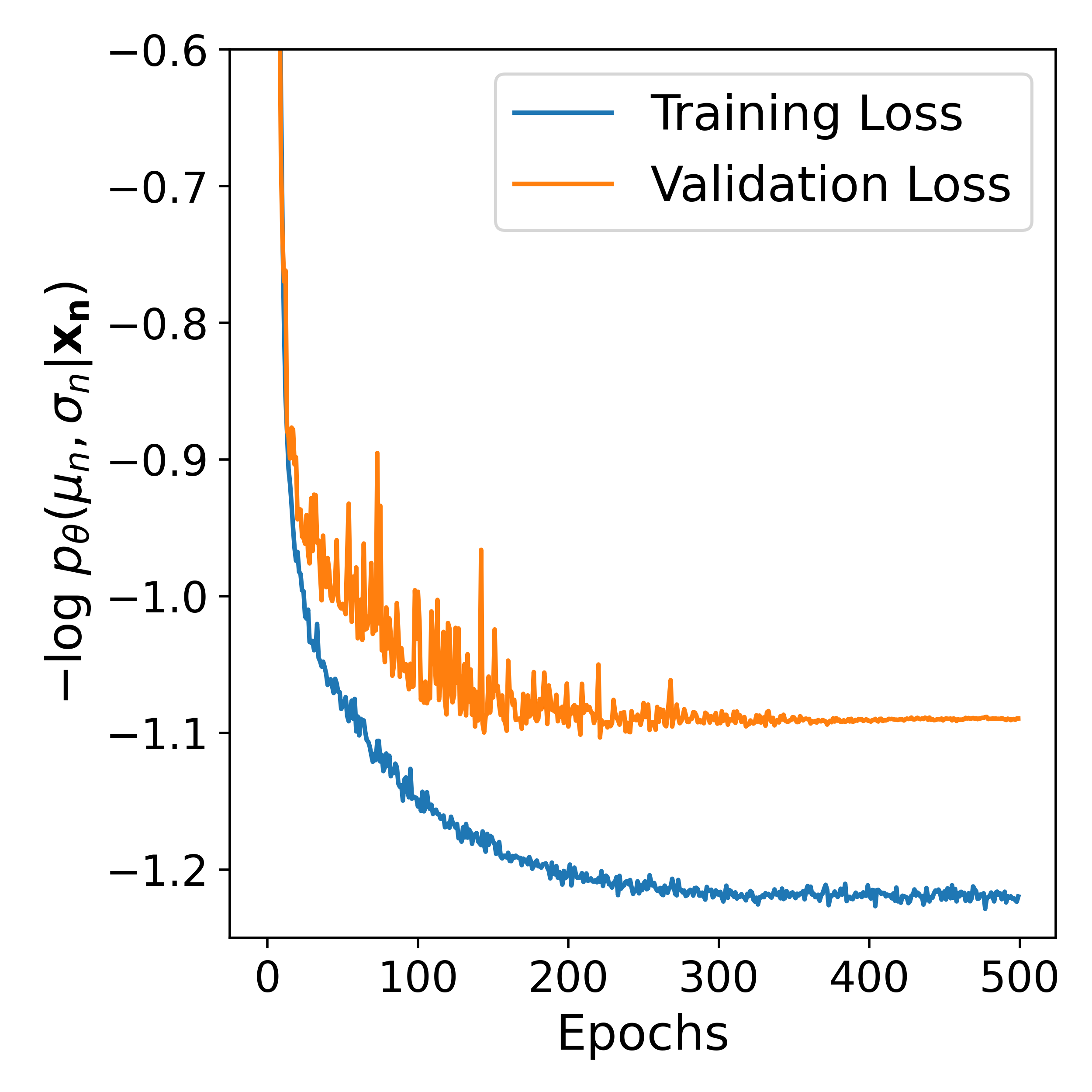}
\end{center}
\caption{{\bf Left:} Neural network architecture of our best performing model. We used a batch size of $100$ and Adam as an optimizer with a starting learning rate of $10^{-4}$. {\bf Right:} The training behavior of our network.}
\label{fig:hyperparameters} 
\end{figure*}

As a proof of concept, we utilize a standard architecture using convolutional and pooling layers, followed by at least one dense layer. We provide the network with information about the source's redshift at the first of these dense layers. To avoid overfitting, we utilize preprocessing layers, which perform random rotations and flips during training, efficiently augmenting the training dataset. We found that this significantly improved performance. We leave a discussion on how other types of architectures, for instance, using geometric deep learning ~\citep[see][]{2021arXiv210413478B}, affect the performance for the future.\footnote{In this work, we utilize Keras~\citep{chollet2015keras} and Tensorflow~\citep{tensorflow2015-whitepaper} for our experiments.}

To enable stable training for a large variety of hyperparameters, we first trained our networks using several standard regression loss functions (e.g.~mean squared error loss of the logarithmic mass). A few hundred epochs of training are typically sufficient. To assess the performance beyond the values for the losses, we check the scatter of masses on the respective training and validation sets for additional biases. More details on our choices and associated scans can be found in Appendix~\ref{app:hyperparameter}.

In our hyperparameter scan, we identify several promising architectures and perform further analysis of these architectures. In particular, we train our network from scratch using a negative log-likelihood for each data sample to predict the mean and standard deviation of a Gaussian:
\begin{equation}
-\log\,p_\theta(y_n|\mathbf{x}_n) = \frac{\log\left(\sigma^2_\theta(\mathbf{x_n})\right)}{2}+\frac{\left( y_n-\mu_\theta(\mathbf{x_n}) \right)^2}{2\sigma^2_\theta(\mathbf{x_n})} + \text{constant}\,,
\label{eq:loss}
\end{equation}
where ${\bf x}_n$ denotes the data, $y_n$ the data label, $\sigma_\theta,~\mu_\theta$ are our predictions which depend on the neural network parameters $\theta.$ The relevant hyperparameters and the training curve for one of our well-performing models can be found in Figure~\ref{fig:hyperparameters}.

Finally, to address the systematic uncertainty in the neural network prediction, we opt for an ensemble method as presented in~\citep{lakshminarayanan1612simple} and leave Bayesian approaches for the future \citep{2015arXiv150602142G}. In practice, we repeat the training procedure with $N$ random weight initializations and the final predictions are calculated via:
\begin{eqnarray}
\log{\overline{M}_{500c}^{\rm NN}}&=&\frac{1}{N}\sum_{i=1}^N \mu_{\theta_i}(x)~,\\
\sigma_*^2(x)&=&\frac{1}{N}\sum_{i=1}^N \sigma^2_{\theta_i}(x)+\mu^2_{\theta_i}(x)-\mu^2_{*}(x)~.
\end{eqnarray}
We now turn to a discussion of the results obtained using this approach for mass estimation.

\section{Results}
\label{sec:results}
\begin{figure*}[t]
\centering
\includegraphics[width=0.73\textwidth]{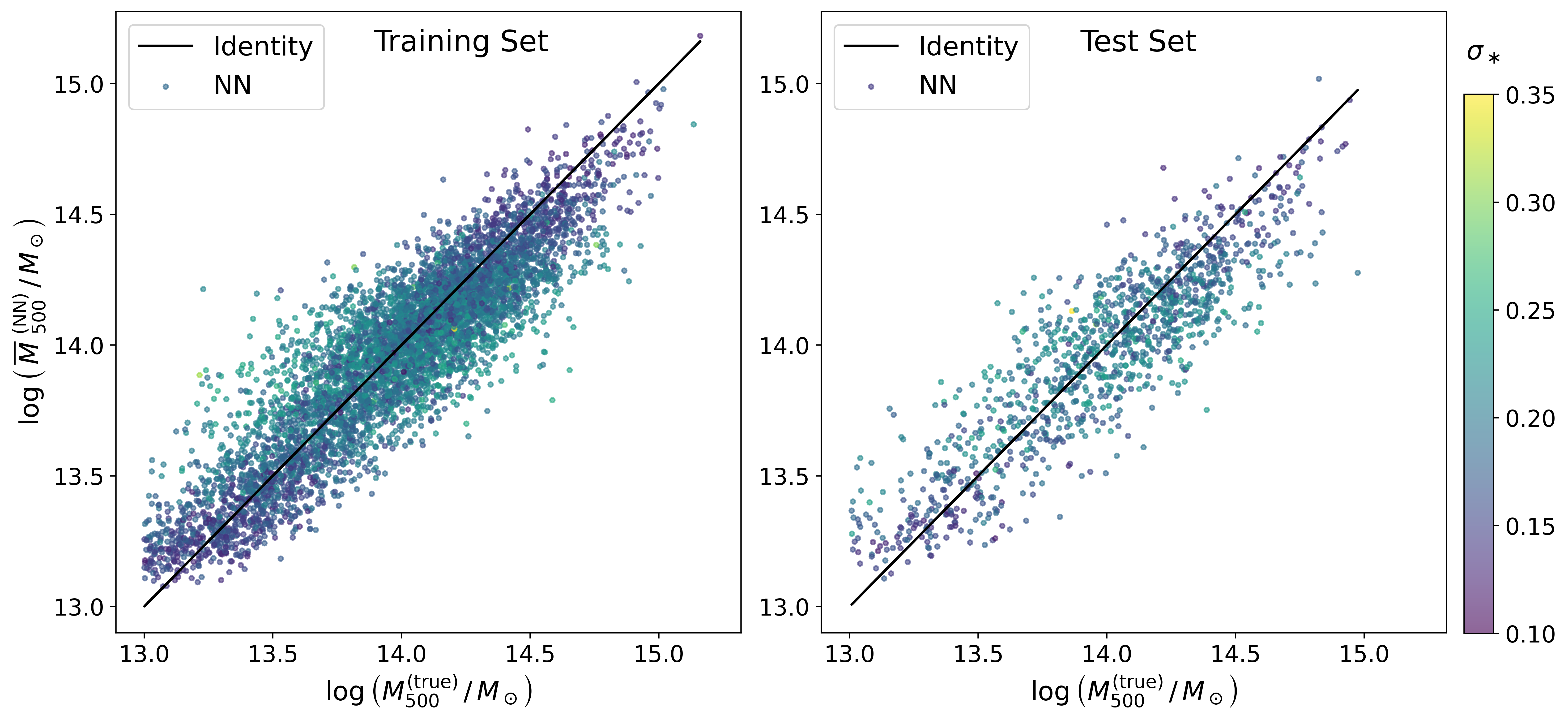}
\includegraphics[width=0.26\textwidth]{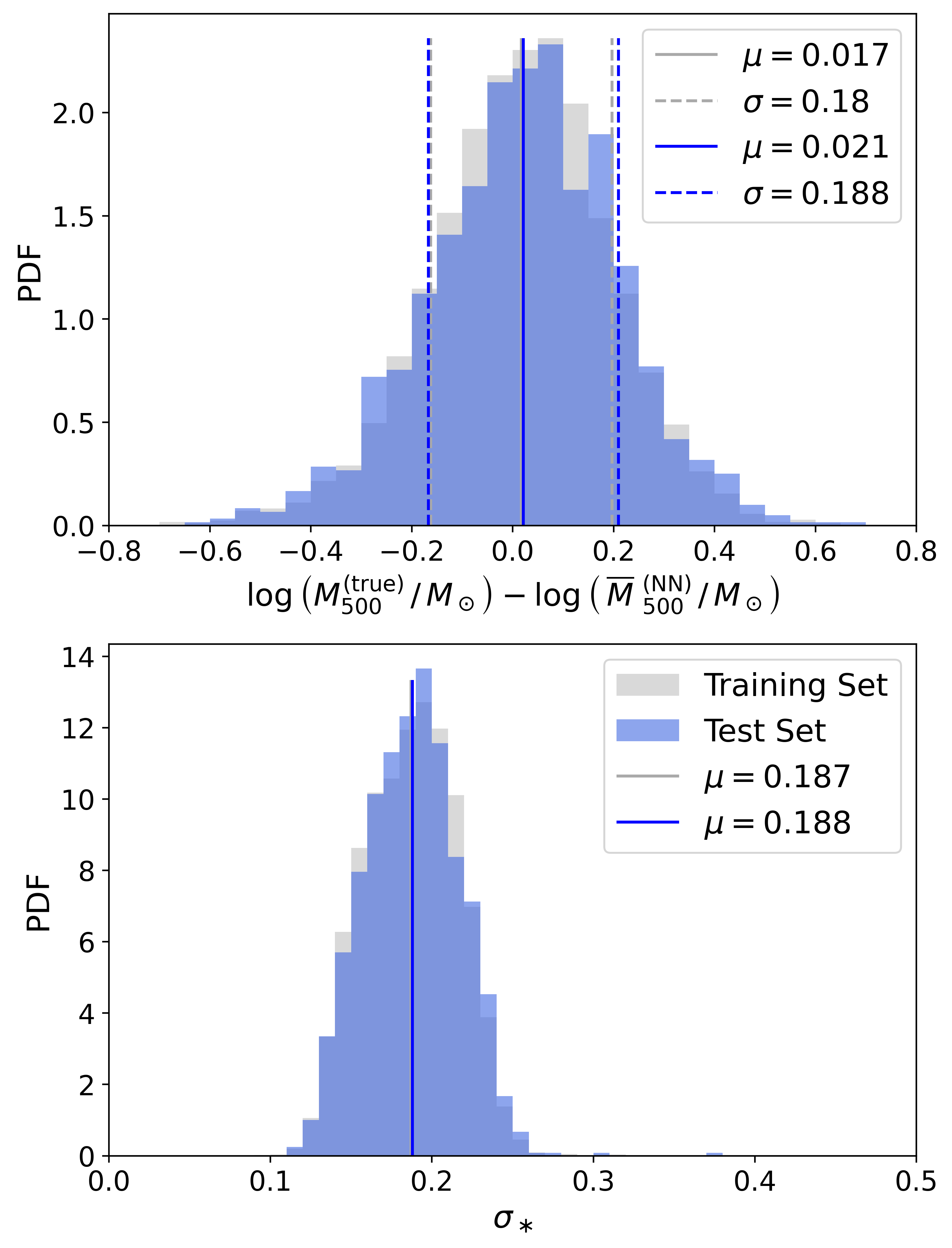}
 \caption{{\bf NN on simulations:} Overview of mass estimation on eFEDS simulations using our ensemble of $30$ convolutional neural networks trained with our likelihood loss from Eq.~\eqref{eq:loss} (cf.~Figure~\ref{fig:hyperparameters} for the hyperparameters). 
{\bf Left:} The mass scatter between predicted mean masses and masses from the simulations on the training set. The colors indicate our predicted standard deviation. {\bf Middle:} Our mean mass predictions on the test set. {\bf Right top:} The distribution of our error $\log_{10}{\left(\mu_*/M_{\odot}\right)}-\log_{10}{\left(M^{\rm true}_{500c}/M_{\odot}\right)}.$ {\bf Right bottom:}
 The distribution of our error estimates which show a mean uncertainty of $0.189$ on the test set.}
\label{fig:efeds-predictions}
\end{figure*}
\begin{figure*}[t]
\centering
\includegraphics[width=1\textwidth]{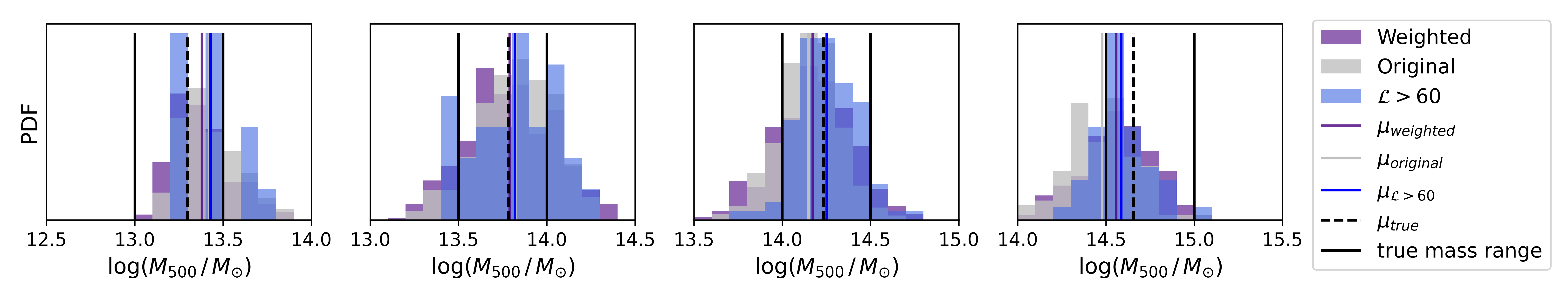}
\caption{{\bf NN biases on simulations:} The mass predictions on eFEDS simulation test sets for different mass ranges to evaluate respective biases between NNs trained on datasets with three different datasets described in the main text.}
\label{fig:efeds-predictions-ranges}
\end{figure*}
\begin{figure*}
\begin{center}
\includegraphics[width=0.5\textwidth]{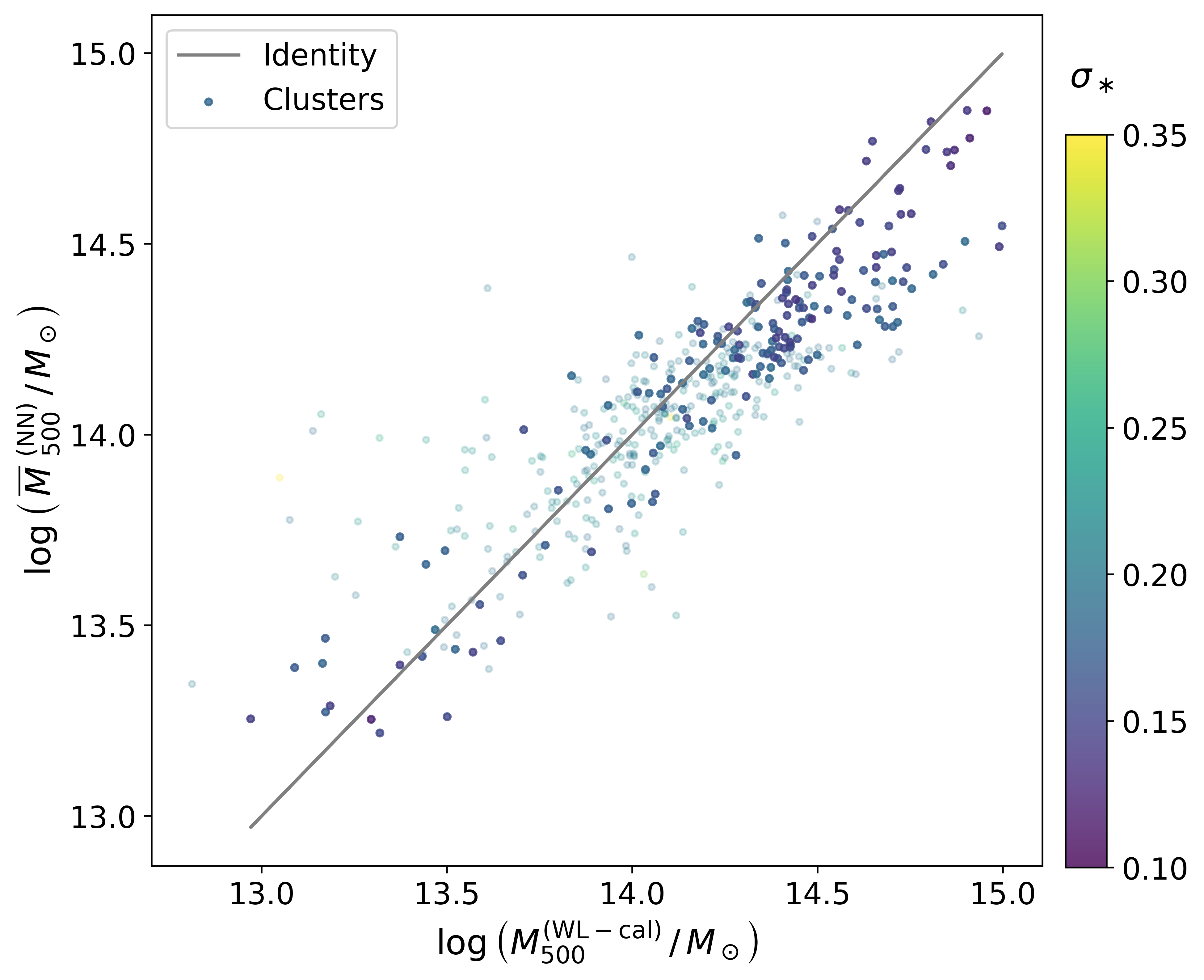}
\includegraphics[width=0.425\textwidth]{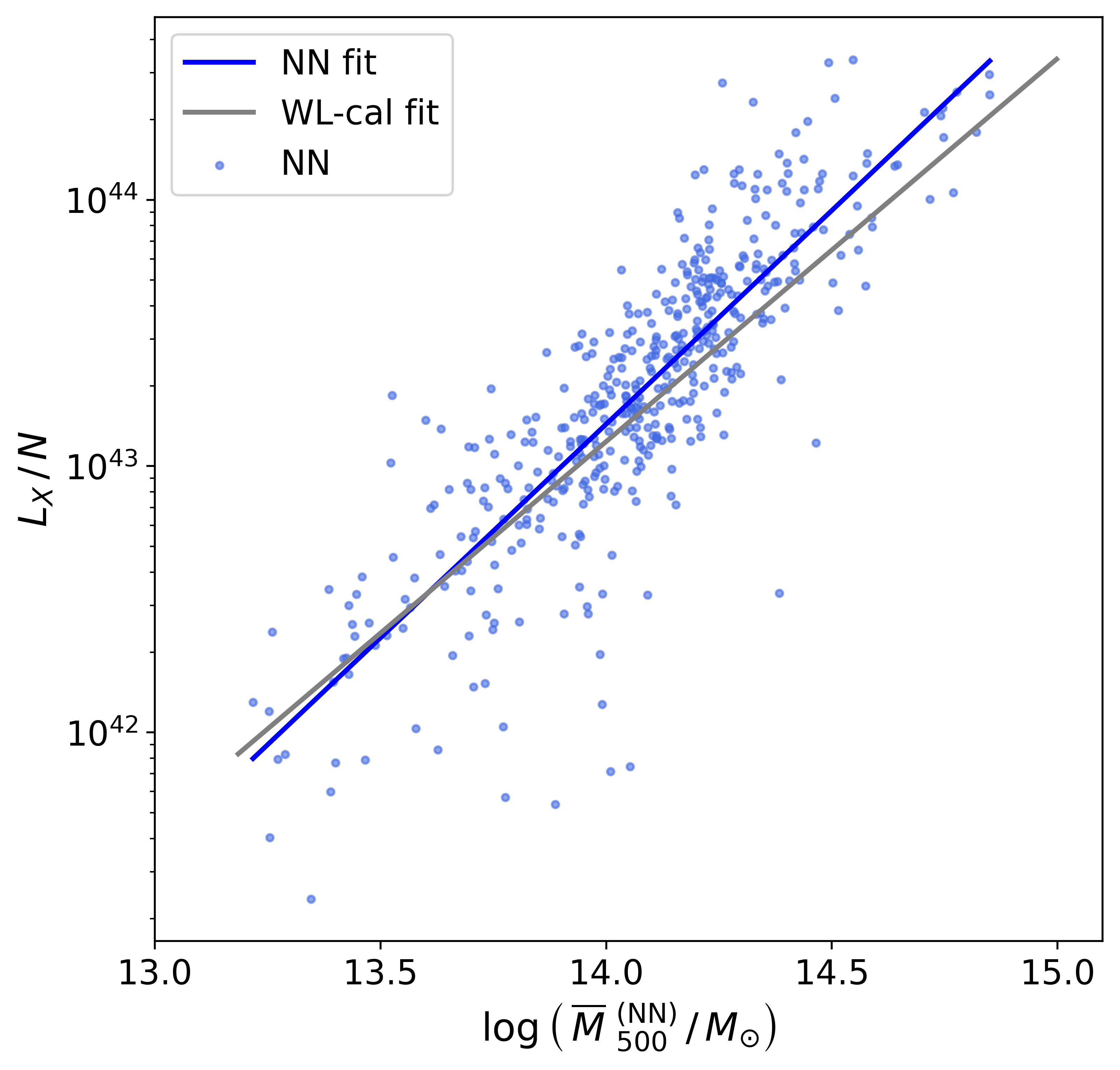}
\end{center}
\caption{{\bf NN on eFEDS observed data:} Comparison of WL calibrated mass estimates (using luminosity scaling relations as in Eq.~(67) of~\citep{chiu}) and masses obtained from our ensemble neural networks. {\bf Left:} Respective mass predictions on eFEDS clusters. The uncertainties in the mass predictions are color-coded and correspond to the NN uncertainties. {\bf Right:} Correlation between predicted masses on eFEDS observations and the measured luminosities as presented in~\citep{chiu}. 
}
\label{fig:wl-nn-masses}
\end{figure*}

To analyze the performance of our neural networks on eFEDS simulations, we firstly compare the predicted and actual mass distributions in the simulations. As shown in Figure~\ref{fig:efeds-predictions}, predictions on the test set and the scatter follow the ideal slope very closely. We observe a scatter of $\sigma=0.188$ on the test set. Our mean error prediction is identical to this value with a mean error of $\langle\sigma_*\rangle=0.188.$ As shown in Figure~\ref{fig:efeds-predictions-ranges}, we observe a bias for the mass range of $13.0<\log{M_{500}/M_\odot}<13.5$ where we are over-predicting the mass on average. In the mass range $14.5<\log{M_{500}/M_\odot}<15.0$ we are under-predicting the masses respectively. To interpret these biases, we performed two experiments:
\begin{enumerate}
    \item To improve the quality of our training and test sample, we train with a cluster sample which has a detection and extent likelihood larger than $60.$ This reduces the scatter to $\sigma=0.159$ on the test set using the same likelihood cuts. In addition, we see a reduction of the bias for  
    high-mass objects. Clearly this cut reduces the number of available clusters significantly (from $7947$ to $1156$ in total). In addition, it is very encouraging to see that our mean uncertainty also reduces to $\langle\sigma_*\rangle=0.158$.
    \item To change the number of clusters at high and low-mass respectively, we weigh our samples to effectively generate a uniform distribution in mass during training. This ensures, in particular, that the network is more strongly penalized when falsely predicting high-mass clusters. We find that the scatter is slightly increased but we reduce the bias for the high-mass clusters from $-0.177$ to $-0.097$ and for the low-mass clusters from $0.121$ to $0.082$. This is encouraging as we only know approximately the observed distribution of cluster masses and our method should be able to compensate for small differences in the distribution.
\end{enumerate}
These respective scatters in the mass predictions have to be compared with the underlying probabilistic cluster model and the application of scaling relations. First of all, there is the intrinsic scaling relation in the data where in our case the luminosity-mass scaling relation has a scatter of $\sigma=0.2$ (cf.~Figure 6 in~\citet{comparat-simulation}) and the temperature-mass scaling relation which has a $\sigma=0.07$. We see that the scatter in our method depends on the quality of the dataset, i.e.~when selecting clusters with high detection and extent likelihood we reduce the scatter below the luminosity scaling relation. Next, when comparing our scatter with scaling relations, a natural caveat is whether the respective scaling relation provides similar results as a scaling relation which is calibrated on this cluster sample e.g.~using WL observables. To do this we utilize scaling relations which have been calibrated for the eFEDS cluster sample.
When using the cluster luminosities and the luminosity mass scaling relations reported in~\citet{chiu} (cf.~Eq.~67), we recover a scatter of $\sigma=0.197$ on our test dataset with detection likelihood larger than 5 and extent likelihood larger than 6 where we have used the actual luminosities in the simulation. This is comparable to the luminosity scatter in the simulation and $4.8\%$ larger than the scatter we observe for our NN masses. When applying to the higher quality cluster sample, the scatter reduces to $\sigma=0.186$ but is significantly above the NN scatter.  To apply these scaling relations we have used an appropriate selection function based using the cluster luminosity and redshift, although we note only a small effect on the ensemble level. 
 In this analysis, the luminosity is normalized with the factor

\begin{equation}
N = \left[ \frac{M_{500}}{M_{\rm piv}} \right]^{\left( \delta_{L_X}\!\ln\left[\frac{1+z}{(1+z_{\rm piv}}\right]\right)} \left[ \frac{E(z)}{E(z_{\rm piv})} \right]^{C_{SS,L_X}} \left[ \frac{1+z}{1+z_{\rm piv}} \right]^{\gamma_{L_X}} ,
\end{equation}
 where the evolution factor $E(z)=H(z)/H_0$, the pivotal mass $M_{\rm piv}=1.4\cdot 10^{14}M_\odot$ and the pivotal redshift $z_{\rm piv}=0.35$. Moreover, the scaling relation parameters as calibrated in the analysis are $\delta_{L_X}=-0.07$, $C_{SS,L_X}=2$ and $\gamma_{L_X}=-0.51$. For this scaling relation analysis, a fiducial flat $\Lambda$CDM cosmology was used with $H_0 = 70$~km~s$^{-1}$~Mpc$^{-1}$, $\Omega_{\mathrm m} =0.3$, $\Omega_{\mathrm b} =0.05$, $\sigma_8 = 0.8$ and $n_s=0.95$.
 
Further, to provide an outlook on inference of cosmological parameters, the scatter is worse when using the count-rate scaling relations on the data with detection likelihood larger than 5 and extent likelihood larger than 6 with the {\it measured} count-rate where we find a scatter of $\sigma=0.265.$ Note that this is without applying the selection function which in light of the effect on the luminosity scaling relation appears to have a small effect in changing the mass predictions for this sample. 
Such a selection function is currently not available for this sample. A more detailed comparison of our systematic uncertainties with systematic uncertainties appearing for the scaling relations between count-rate and weak lensing mass as discussed in~\cite{2021MNRAS.507.5671G} is left for the future.

For both scaling relations we find a significant reduction in scatter. The amount of reduction depends significantly on the data used for training, this does not only dependent on the mass and redshift distribution. 

One further advantage of the NN-based mass estimation is that the training networks use the full morphology information of the input clusters in the X-ray images \citep{Ghirardini2022} compared to other methods and are not impacted by the line-of-sight structure or assumed 3D-morphology of the source when estimating masses \citep{ZuHone2022} or hydrostatic mass bias often a problem for X-ray mass measurements \citep{Scheck2023}.

\begin{figure}[t]
\begin{center}
\includegraphics[width=0.49\textwidth]{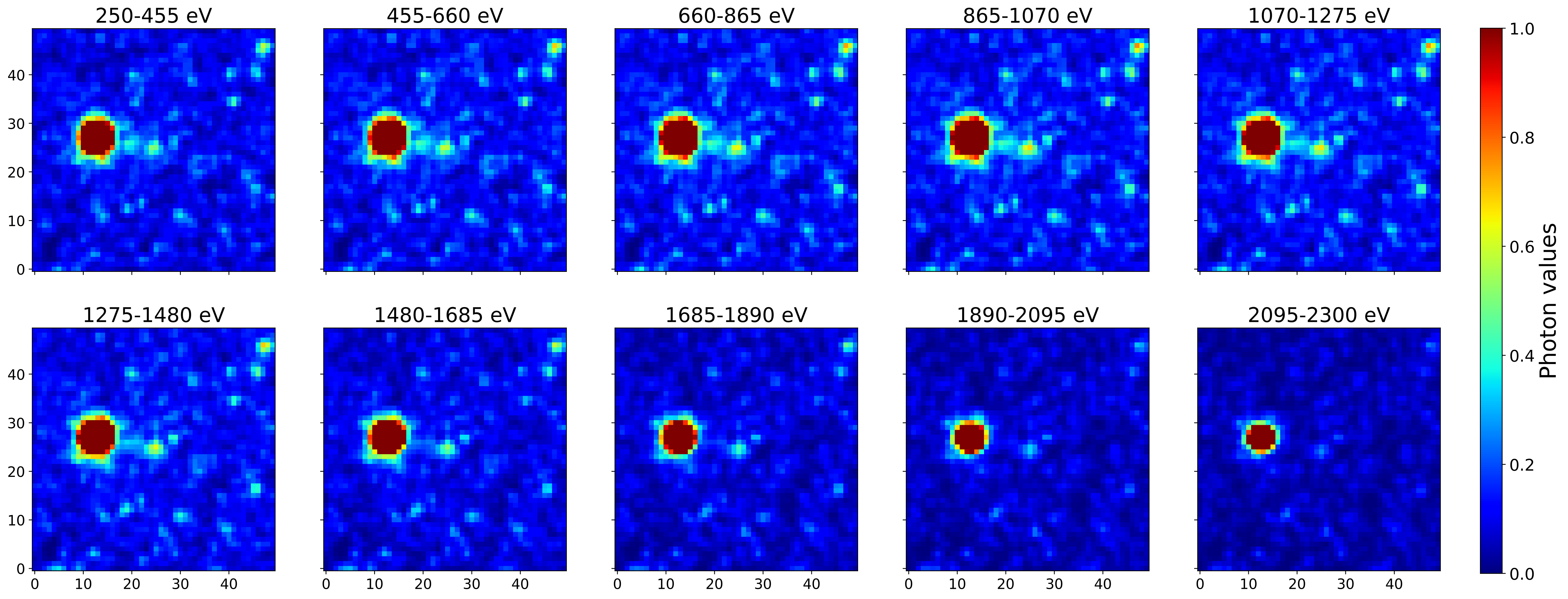}\\[0.2cm]
\includegraphics[width=0.49\textwidth]{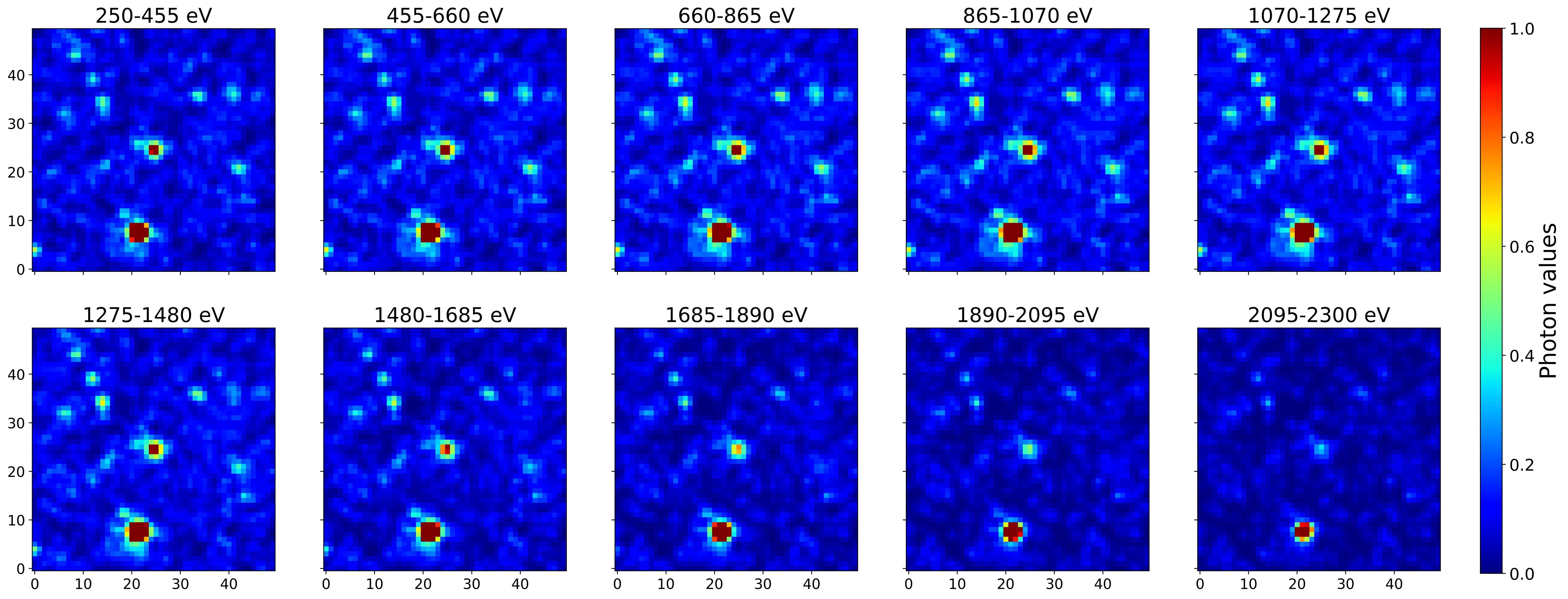}
\end{center}
\caption{All ten channels of the EBI of the galaxy cluster with the highest and lowest $\Delta\mu = \mu_{\rm max}-\mu_{\rm min}$, where $\mu_{\rm max}$ $(\mu_{\rm min})$ corresponds to the highest (smallest) predicted mass by the ensemble members. {\bf Top:}
This cluster -- located as usual at the center of the EBI -- has a mass of $\log{M_{500}/M_\odot}=13.959$ and is at redshift $z=0.211$ in the simulations (Object with SRC\_ID: 10003763 from realization 18). The lowest and highest predicted mean values in our ensemble are $\mu_{\rm min}=13.323$ and $\mu_{\rm max}=14.587$ respectively. For illustration purposes, to make the actual cluster visible, we have clipped the photon values at $1.$ The final mass predictions are $\log{M_{500}/M_{\odot}}=14.079\pm 0.305 .$
{\bf Bottom:} An example of a cluster with little differences among the ensemble (SRC\_ID: 10006556 from realization 9), it has a mass of $\log{M_{500}/M_{\odot}}=13.873$ and is at redshift $z=0.305$. We obtain $\mu_{\rm max}=14.095$ and $\mu_{\rm min}=13.994$ and our final prediction for this cluster is $\log{M_{500}/M_{\odot}}=14.042\pm 0.161 .$
}
\label{fig:delta_mean}
\end{figure}

We note that the predicted means and the respective standard deviations do not vary hugely on an ensemble level. In particular, we observe that the ratio of the individual $\sigma$-values for each network and the correspond ensemble prediction $\sigma_*$ is given as $\langle \sigma/\sigma_*\rangle=0.951\pm 0.039$ where we quote the single standard deviation values and where we have averaged over all clusters in the test sample. On an individual level we report the clusters with the highest and smallest differences
 in the predicted masses (see Figure~\ref{fig:delta_mean}). We often find upon visual inspection that the largest differences in the predicted masses occur when other bright X-ray sources are present in the EBI and our cluster of interest is a less luminous source.

Having seen that our method provides sensible looking mass estimates on eFEDS simulations, we now estimate the masses for the eFEDS cluster sample with extent likelihood larger than 6 and detection likelihood larger than 5. We use the ensemble of neural networks which we have trained on data with the same selection criteria. We show the scatter between our NN predicted masses and the masses obtained using WL-calibrated luminosity scaling relations in Figure~\ref{fig:wl-nn-masses}. We observe that both predictions agree for clusters where our NN ensemble predicts a low uncertainty $\sigma_\ast<0.185$ (more visible points correspond to clusters with such a low uncertainty).  We note that for the few clusters present in the eFEDS cluster sample which have a mass below the range we have trained on, our neural network ensemble still predicts masses in the mass regime it was trained on and does not generalize for this data outside of the known regime.

\begin{figure*}[t]
\begin{center}
\includegraphics[width=0.74\textwidth]{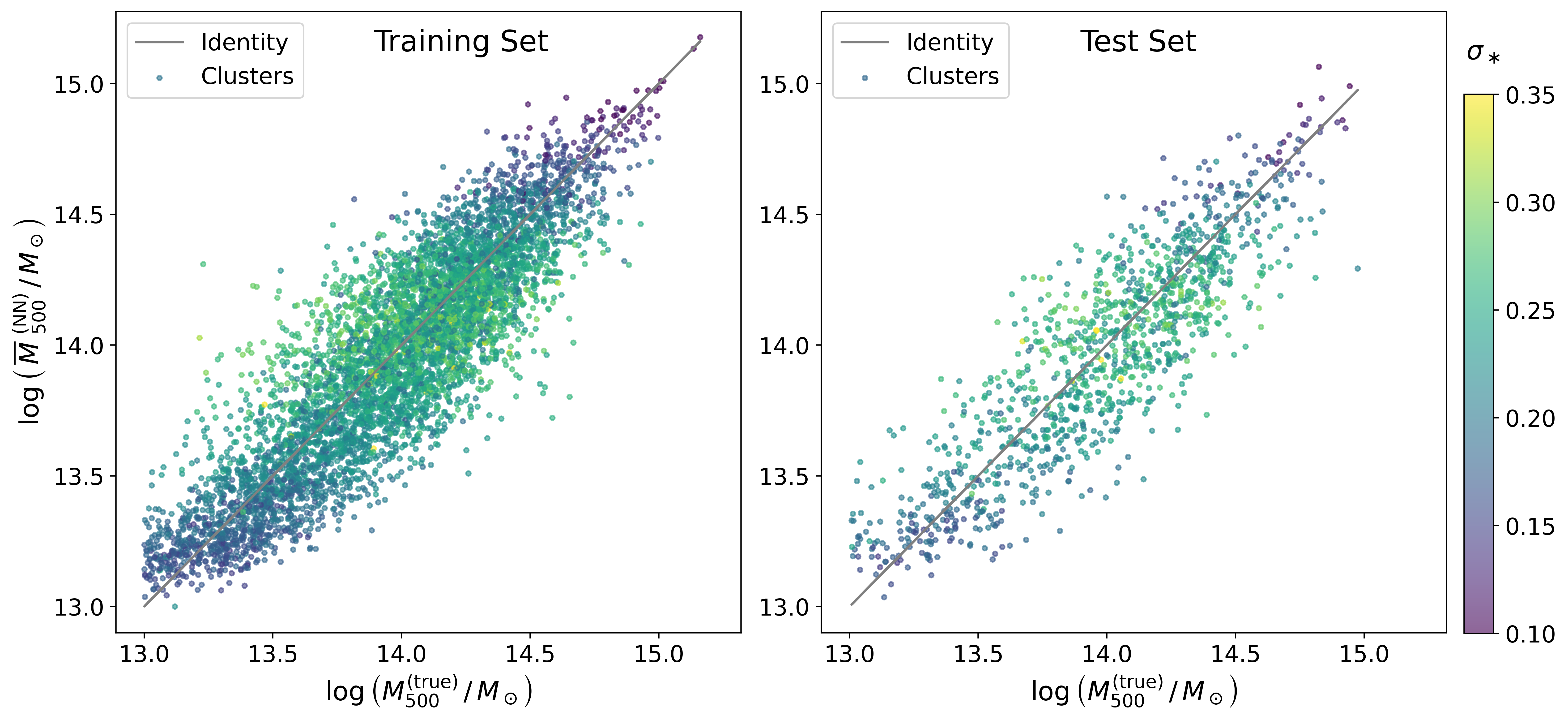}
\includegraphics[width=0.24\textwidth]{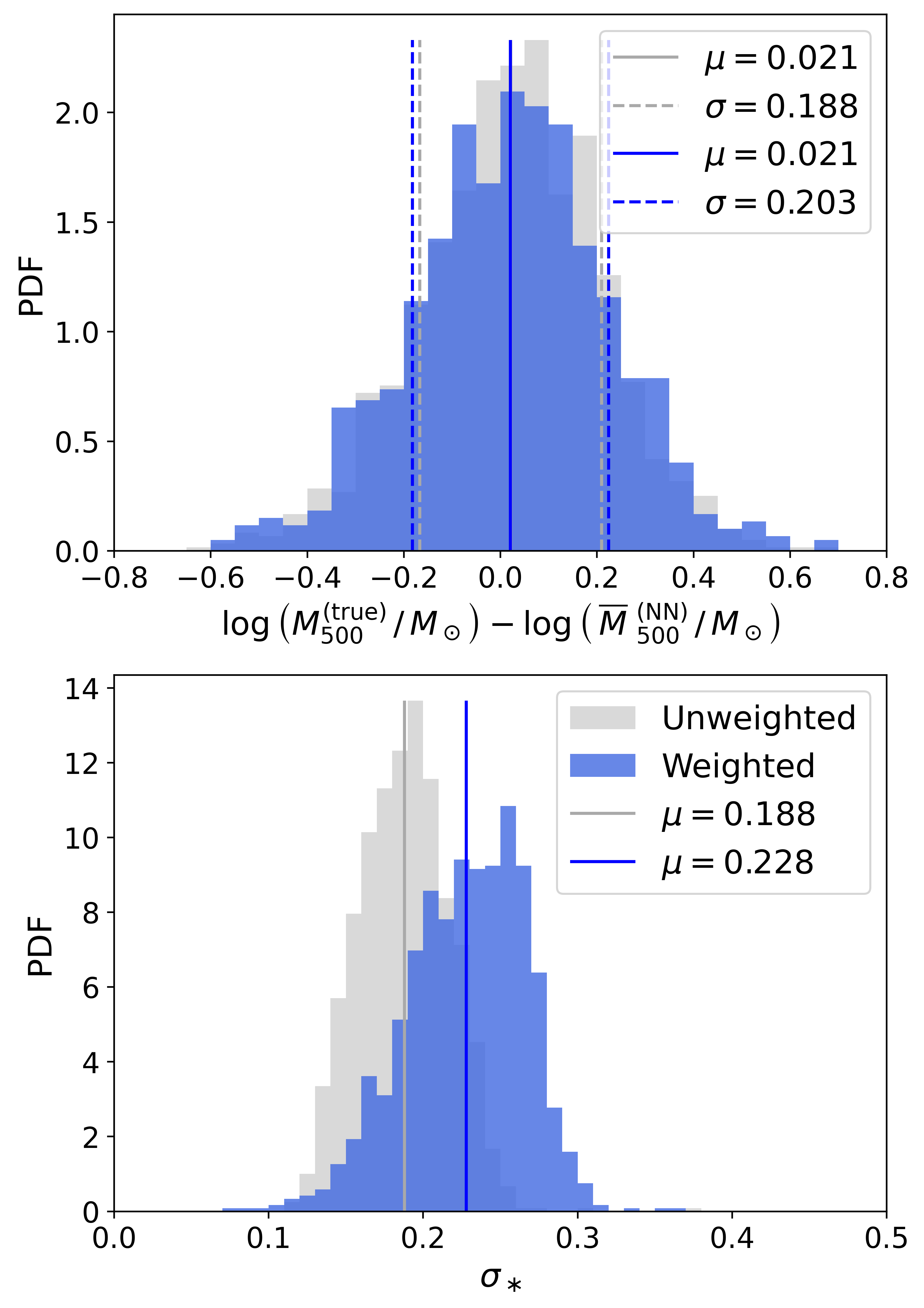}
\end{center}
\caption{{\bf NN trained on uniform cluster sample:} Overview of mass estimation on eFEDS simulations using our deep ensemble trained with class weights such that the effective mass distribution is close to uniform, illustrating the robustness to uncertainty in the underlying mass distribution.
{\bf Left:} The mass scatter between predicted means and masses from the simulations on the training set. The colors indicate our predicted standard deviation. {\bf Middle:} Our mass predictions on the test set. {\bf Right:} Comparison of the residuals and distribution of standard deviations in our test set mass distributions between our normal training procedure with masses distributed as shown in Fig.~\ref{fig:distributions} and our weighted clusters. 
 The distribution of our error estimates shows a mean uncertainty of $0.228$.}
 \label{fig:weights}
\end{figure*}

Further, to compare our predicted masses with the luminosity-mass estimates of the clusters, we show the scatter between the luminosity and our mass estimates on the right of Figure~\ref{fig:wl-nn-masses}. Overall, we find a linear relation which is close to the slope identified via the WL-calibrated scaling relation. However, we find deviations from the WL-calibrated masses at high masses. 
Further analysis of the features being used by the NN ensemble, ultimately aiming at a data-driven scaling relation, is beyond the scope of this paper.

\section{Conclusions}
\label{sec:conclusions}
We have demonstrated that galaxy cluster masses can be estimated using NN ensemble predictions when applied to the eFEDS-field of {\it eROSITA}, both to the respective simulations and actual observations. Depending on the training data, we observe a significant reduction in scatter in comparison to luminosity based scaling relations from $\sigma=0.186$ to $\sigma=0.159$ on a sample with higher detection and extent likelihood and from $\sigma=0.197$ to $\sigma=0.188$ on the entire sample. Compared to count-rate based scaling relations, the improvement is from $\sigma=0.265$ to $\sigma=0.188$.
Our approach is applicable to clusters at different redshifts and we are not required to remove other clusters or point sources from the respective images to mimic a realistic observational set-up. 
Going beyond existing NN methods for cluster mass estimation, our method provides uncertainty measurements of the NN predicted masses for each cluster. Our ML approach can be integrated into a highly developed workflow for estimating cluster masses and their subsequent use for cosmological parameter inference. 
The interplay with each of these components is important to understand shortcomings and potentials for improved mass estimates in the future:

\begin{itemize}
    \item {\bf WL and other additional measurements:} Given the dependence on our simulations of eFEDS clusters, our NN methods do not require in addition WL measurements for (a subset of) the X-ray selected cluster sample. Any constraints, e.g. for a subsequent cosmological analysis, arising from the requirement of availability of WL information can be circumvented.

    Our model can be easily expanded and improved by adding new features and observations, similar to using redshift information to the network. For instance,  richness information coming from optical observations of clusters of galaxies, as being developed in the context of {\it Euclid}, promises to improve the NN-predictions in the high-mass end \citep[e.g.][]{EuclidCollaboration2022arXiv220614944E}. As in any other model, adding new multi-wavelength data requires appropriate calibration and will be used in future work.
   
    \item {\bf Simulations:} as our ML approach is obtained from the underlying data model, it heavily depends on the data used for training. To make our method work, it is crucial that the training data is of sufficient quality. This requires, for instance, that the training data contains clusters in the appropriate mass regime and that the clusters in the training sample are ideally very close to the cluster sample the method is applied on. At this stage, a generalization beyond properties captured by the training data is not guaranteed. Throughout this project, we often encountered performance deterioration when including different cluster samples for training. Addressing the independence of the training data is a clear future goal but -- as demonstrated here -- can be circumvented by utilizing a dedicated training set. Implicitly, our method depends on the data used to shape the simulations and in particular on the underlying scaling relations. 
    However, we crucially observe that the mass distribution of clusters in the training sample is not of high importance as showcased when using a uniform distribution of masses (Figure \ref{fig:weights}). This is particularly encouraging as this allows for generalization across different mass distributions from different cosmologies.
  
    \item {\bf ML vs.~known astrophysical features:} there are two approaches for predicting masses using ML; either using known astrophysical features, e.g.~the measured luminosities or count-rate, as the input (cf.~\citet{Green_2019}) or directly the photon information. Here, we explore the latter and demonstrate that it provides competitive mass estimates. Future studies will provide more information on which method ultimately predicts the most accurate mass estimates. It would be very interesting to compare the ML features with previously identified features (e.g.~using, appropriate dimensional reduction and symbolic regression, see~\citet{2201.01305} for work in this direction).
    \end{itemize}

Finally, we summarize the advances from the ML-based mass predictions, presented here:

\begin{itemize}
    \item We demonstrate, for the first time, that meaningful uncertainty measures can be provided with the mass estimates in X-ray cluster mass estimations with ML and, in particular, neural networks.  This is a crucial requirement for integrating ML-based methods into cosmological analyses with cluster counts.

    \item As our simulations also includes clusters with masses as low as $10^{13}M_{\odot},$ we are able to demonstrate for the first time that this neural network approach to X-ray cluster mass estimates also works in this mass regime without introducing large biases. A further successful extension to the low-mass regime would be very interesting and could dramatically increase the sample utilized for cosmology, we found that objects with a high detection and extent likelihood provide an avenue forward.

    \item Instead of a single channel that can only capture information about the total number of photons, we utilize an input format that also captures the energy information of photons. Our EBIs enable the neural networks, at least in principle, to utilize energy-dependent information such as the cluster temperature.
\end{itemize}

One of the immediate next objectives is to apply our method to other {\it eROSITA} cluster samples, particularly the upcoming All-Sky Survey data. To make our method applicable to these observations, we need to ensure that the performance does not decrease due to the different exposure times for individual clusters in those samples, as the first All-Sky survey data is shallower than the eFEDS data used in this work. A further extension to observations of other X-ray telescopes (e.g.~XMM, Chandra), despite being very interesting, would require dedicated datasets to train the ML method appropriately.

As this paper was in its final stage, the preprint~\citet{2023arXiv230300005H}, which discusses a similar question, appeared on astro-ph. Our approach differs from \citet{2023arXiv230300005H} by the use of the simulation data sets for training. The sample of simulated clusters used in this work represents the {\it eROSITA} cluster selection. Our method could hence be successfully applied to the {\it eROSITA} survey observations and compared with the observational WL mass measurements utlized for the same sample, self-consistently. Through this work, we also provide a clear path toward using ML-based masses in cosmological analyses.  Additionally, we successfully utilize a likelihood loss for the first time, enabling uncertainty estimates, i.e., a prerequisite for employing ML-based masses in future scaling relations and cosmology analyses.

\section*{Acknowledgments}
 This work is based on data from {\it eROSITA}, the soft X-ray instrument aboard SRG, a joint Russian-German science mission supported by the Russian Space Agency (Roskosmos), in the interests of the Russian Academy of Sciences represented by its Space Research Institute (IKI), and the Deutsches Zentrum f{\"{u}}r Luft und Raumfahrt (DLR). The SRG spacecraft was built by Lavochkin Association (NPOL) and its subcontractors and is operated by NPOL with support from the Max Planck Institute for Extraterrestrial Physics (MPE).

The development and construction of the {\it eROSITA} X-ray instrument was led by MPE, with contributions from the Dr. Karl Remeis Observatory Bamberg \& ECAP (FAU Erlangen-Nuernberg), the University of Hamburg Observatory, the Leibniz Institute for Astrophysics Potsdam (AIP), and the Institute for Astronomy and Astrophysics of the University of T{\"{u}}bingen, with the support of DLR and the Max Planck Society. The Argelander Institute for Astronomy of the University of Bonn and the Ludwig Maximilians Universit{\"{a}}t Munich also participated in the science preparation for {\it eROSITA}.

The {\it eROSITA} data shown here were processed using the {\tt eSASS} software system developed by the German {\it eROSITA} consortium.
\\

E.B., A.L., V.G., X.Z., and C.G. acknowledge financial support from the European Research Council (ERC) Consolidator Grant under the European Union’s Horizon 2020 research and innovation programme (grant agreement CoG DarkQuest No 101002585). 

\section*{Software packages used}
We used the following packages: Numpy \citep{harris2020array}, Tensorflow \citep{tensorflow2015-whitepaper}, Astropy \citep{astropy:2022}, Pandas \citep{mckinney-proc-scipy-2010}, Matplotlib \citep{Hunter:2007}, Scipy \citep{2020SciPy-NMeth}, Scikit-learn \citep{scikit-learn}.

\bibliography{references}

\begin{thebibliography}{52}
\expandafter\ifx\csname natexlab\endcsname\relax\def\natexlab#1{#1}\fi

\bibitem[{Abadi {et~al.}(2015)Abadi, Agarwal, Barham, Brevdo, Chen, Citro,
  Corrado, Davis, Dean, Devin, Ghemawat, Goodfellow, Harp, Irving, Isard, Jia,
  Jozefowicz, Kaiser, Kudlur, Levenberg, Man\'{e}, Monga, Moore, Murray, Olah,
  Schuster, Shlens, Steiner, Sutskever, Talwar, Tucker, Vanhoucke, Vasudevan,
  Vi\'{e}gas, Vinyals, Warden, Wattenberg, Wicke, Yu, \&
  Zheng}]{tensorflow2015-whitepaper}
Abadi, M., Agarwal, A., Barham, P., {et~al.} 2015, {TensorFlow}: Large-Scale
  Machine Learning on Heterogeneous Systems, software available from
  tensorflow.org

\bibitem[{{Astropy Collaboration} {et~al.}(2022){Astropy Collaboration},
  {Price-Whelan}, {Lim}, {Earl}, {Starkman}, {Bradley}, {Shupe}, {Patil},
  {Corrales}, {Brasseur}, {N{"o}the}, {Donath}, {Tollerud}, {Morris},
  {Ginsburg}, {Vaher}, {Weaver}, {Tocknell}, {Jamieson}, {van Kerkwijk},
  {Robitaille}, {Merry}, {Bachetti}, {G{"u}nther}, {Aldcroft},
  {Alvarado-Montes}, {Archibald}, {B{'o}di}, {Bapat}, {Barentsen}, {Baz{'a}n},
  {Biswas}, {Boquien}, {Burke}, {Cara}, {Cara}, {Conroy}, {Conseil}, {Craig},
  {Cross}, {Cruz}, {D'Eugenio}, {Dencheva}, {Devillepoix}, {Dietrich},
  {Eigenbrot}, {Erben}, {Ferreira}, {Foreman-Mackey}, {Fox}, {Freij}, {Garg},
  {Geda}, {Glattly}, {Gondhalekar}, {Gordon}, {Grant}, {Greenfield}, {Groener},
  {Guest}, {Gurovich}, {Handberg}, {Hart}, {Hatfield-Dodds}, {Homeier},
  {Hosseinzadeh}, {Jenness}, {Jones}, {Joseph}, {Kalmbach}, {Karamehmetoglu},
  {Ka{l}uszy{'n}ski}, {Kelley}, {Kern}, {Kerzendorf}, {Koch}, {Kulumani},
  {Lee}, {Ly}, {Ma}, {MacBride}, {Maljaars}, {Muna}, {Murphy}, {Norman},
  {O'Steen}, {Oman}, {Pacifici}, {Pascual}, {Pascual-Granado}, {Patil},
  {Perren}, {Pickering}, {Rastogi}, {Roulston}, {Ryan}, {Rykoff}, {Sabater},
  {Sakurikar}, {Salgado}, {Sanghi}, {Saunders}, {Savchenko}, {Schwardt},
  {Seifert-Eckert}, {Shih}, {Jain}, {Shukla}, {Sick}, {Simpson},
  {Singanamalla}, {Singer}, {Singhal}, {Sinha}, {Sip{H{o}}cz}, {Spitler},
  {Stansby}, {Streicher}, {{{S}}umak}, {Swinbank}, {Taranu}, {Tewary},
  {Tremblay}, {Val-Borro}, {Van Kooten}, {Vasovi{'c}}, {Verma}, {de Miranda
  Cardoso}, {Williams}, {Wilson}, {Winkel}, {Wood-Vasey}, {Xue}, {Yoachim},
  {Zhang}, {Zonca}, \& {Astropy Project Contributors}}]{astropy:2022}
{Astropy Collaboration}, {Price-Whelan}, A.~M., {Lim}, P.~L., {et~al.} 2022,
  apj, 935, 167

\bibitem[{{Bahar} {et~al.}(2022){Bahar}, {Bulbul}, {Clerc}, {Ghirardini},
  {Liu}, {Nandra}, {Pacaud}, {Chiu}, {Comparat}, {Ider-Chitham}, {Klein},
  {Liu}, {Merloni}, {Migkas}, {Okabe}, {Ramos-Ceja}, {Reiprich}, {Sanders}, \&
  {Schrabback}}]{Bahar2022}
{Bahar}, Y.~E., {Bulbul}, E., {Clerc}, N., {et~al.} 2022, \aap, 661, A7

\bibitem[{{Bocquet} {et~al.}(2019){Bocquet}, {Dietrich}, {Schrabback}, {Bleem},
  {Klein}, {Allen}, {Applegate}, {Ashby}, {Bautz}, {Bayliss}, {Benson},
  {Brodwin}, {Bulbul}, {Canning}, {Capasso}, {Carlstrom}, {Chang}, {Chiu},
  {Cho}, {Clocchiatti}, {Crawford}, {Crites}, {de Haan}, {Desai}, {Dobbs},
  {Foley}, {Forman}, {Garmire}, {George}, {Gladders}, {Gonzalez}, {Grandis},
  {Gupta}, {Halverson}, {Hlavacek-Larrondo}, {Hoekstra}, {Holder}, {Holzapfel},
  {Hou}, {Hrubes}, {Huang}, {Jones}, {Khullar}, {Knox}, {Kraft}, {Lee}, {von
  der Linden}, {Luong-Van}, {Mantz}, {Marrone}, {McDonald}, {McMahon}, {Meyer},
  {Mocanu}, {Mohr}, {Morris}, {Padin}, {Patil}, {Pryke}, {Rapetti},
  {Reichardt}, {Rest}, {Ruhl}, {Saliwanchik}, {Saro}, {Sayre}, {Schaffer},
  {Shirokoff}, {Stalder}, {Stanford}, {Staniszewski}, {Stark}, {Story},
  {Strazzullo}, {Stubbs}, {Vanderlinde}, {Vieira}, {Vikhlinin}, {Williamson},
  \& {Zenteno}}]{Bocquet2019}
{Bocquet}, S., {Dietrich}, J.~P., {Schrabback}, T., {et~al.} 2019, \apj, 878,
  55

\bibitem[{{Bronstein} {et~al.}(2021){Bronstein}, {Bruna}, {Cohen}, \&
  {Veli{\v{c}}kovi{\'c}}}]{2021arXiv210413478B}
{Bronstein}, M.~M., {Bruna}, J., {Cohen}, T., \& {Veli{\v{c}}kovi{\'c}}, P.
  2021, arXiv e-prints, arXiv:2104.13478

\bibitem[{{Brunner} {et~al.}(2022){Brunner}, {Liu}, {Lamer}, {Georgakakis},
  {Merloni}, {Brusa}, {Bulbul}, {Dennerl}, {Friedrich}, {Liu}, {Maitra},
  {Nandra}, {Ramos-Ceja}, {Sanders}, {Stewart}, {Boller}, {Buchner}, {Clerc},
  {Comparat}, {Dwelly}, {Eckert}, {Finoguenov}, {Freyberg}, {Ghirardini},
  {Gueguen}, {Haberl}, {Kreykenbohm}, {Krumpe}, {Osterhage}, {Pacaud},
  {Predehl}, {Reiprich}, {Robrade}, {Salvato}, {Santangelo}, {Schrabback},
  {Schwope}, \& {Wilms}}]{Brunner2022}
{Brunner}, H., {Liu}, T., {Lamer}, G., {et~al.} 2022, \aap, 661, A1

\bibitem[{{Bulbul} {et~al.}(2019){Bulbul}, {Chiu}, {Mohr}, {McDonald},
  {Benson}, {Bautz}, {Bayliss}, {Bleem}, {Brodwin}, {Bocquet}, {Capasso},
  {Dietrich}, {Forman}, {Hlavacek-Larrondo}, {Holzapfel}, {Khullar}, {Klein},
  {Kraft}, {Miller}, {Reichardt}, {Saro}, {Sharon}, {Stalder}, {Schrabback}, \&
  {Stanford}}]{Bulbul2019}
{Bulbul}, E., {Chiu}, I.~N., {Mohr}, J.~J., {et~al.} 2019, \apj, 871, 50

\bibitem[{{Bulbul} {et~al.}(2022){Bulbul}, {Liu}, {Pasini}, {Comparat},
  {Hoang}, {Klein}, {Ghirardini}, {Salvato}, {Merloni}, {Seppi}, {Wolf},
  {Anderson}, {Bahar}, {Brusa}, {Br{\"u}ggen}, {Buchner}, {Dwelly},
  {Ibarra-Medel}, {Ider Chitham}, {Liu}, {Nandra}, {Ramos-Ceja}, {Sanders}, \&
  {Shen}}]{Bulbul2022}
{Bulbul}, E., {Liu}, A., {Pasini}, T., {et~al.} 2022, \aap, 661, A10

\bibitem[{{Chiu} {et~al.}(2022){Chiu}, {Ghirardini}, {Liu}, {Grandis},
  {Bulbul}, {Bahar}, {Comparat}, {Bocquet}, {Clerc}, {Klein}, {Liu}, {Li},
  {Miyatake}, {Mohr}, {More}, {Oguri}, {Okabe}, {Pacaud}, {Ramos-Ceja},
  {Reiprich}, {Schrabback}, \& {Umetsu}}]{chiu}
{Chiu}, I.~N., {Ghirardini}, V., {Liu}, A., {et~al.} 2022, \aap, 661, A11

\bibitem[{Chollet {et~al.}(2015)}]{chollet2015keras}
Chollet, F. {et~al.} 2015, Keras, \url{https://keras.io}

\bibitem[{{Cohn} \& {Battaglia}(2020)}]{Cohn_2019}
{Cohn}, J.~D. \& {Battaglia}, N. 2020, \mnras, 491, 1575

\bibitem[{{Comparat} {et~al.}(2020){Comparat}, {Eckert}, {Finoguenov},
  {Schmidt}, {Sanders}, {Nagai}, {Lau}, {K{\"a}}, {fer}, {Pacaud}, {Clerc},
  {Reiprich}, {Bulbul}, {Chitham}, {Chiang}, {Ghirardini}, {Gonzalez-Perez},
  {Gozaliasl}, {Fitzpatrick}, {Klypin}, {Merloni}, {Nandra}, {Liu}, {Prada},
  {Ramos-Ceja}, {Salvato}, {Seppi}, {Tempel}, \& {Yepes}}]{comparat-simulation}
{Comparat}, J., {Eckert}, D., {Finoguenov}, A., {et~al.} 2020, The Open Journal
  of Astrophysics, 3, 13

\bibitem[{Comparat {et~al.}(2019)Comparat, Merloni, Salvato, Nandra, Boller,
  Georgakakis, Finoguenov, Dwelly, Buchner, Moro, Clerc, Wang, Zhao, Prada,
  Yepes, Brusa, Krumpe, \& Liu}]{Comparat_2019}
Comparat, J., Merloni, A., Salvato, M., {et~al.} 2019, Monthly Notices of the
  Royal Astronomical Society, 487, 2005

\bibitem[{Dauser {et~al.}(2019)Dauser, Falkner, Lorenz, Kirsch, Peille,
  Cucchetti, Schmid, Brand, Oertel, Smith, {et~al.}}]{dauser2019sixte}
Dauser, T., Falkner, S., Lorenz, M., {et~al.} 2019, Astronomy \& Astrophysics,
  630, A66

\bibitem[{{Euclid Collaboration} {et~al.}(2022){Euclid Collaboration},
  {Bisigello}, {Conselice}, {Baes}, {Bolzonella}, {Brescia}, {Cavuoti},
  {Cucciati}, {Humphrey}, {Hunt}, {Maraston}, {Pozzetti}, {Tortora}, {van
  Mierlo}, {Aghanim}, {Auricchio}, {Baldi}, {Bender}, {Bodendorf}, {Bonino},
  {Branchini}, {Brinchmann}, {Camera}, {Capobianco}, {Carbone}, {Carretero},
  {Castander}, {Castellano}, {Cimatti}, {Congedo}, {Conversi}, {Copin},
  {Corcione}, {Courbin}, {Cropper}, {Da Silva}, {Degaudenzi}, {Douspis},
  {Dubath}, {Duncan}, {Dupac}, {Dusini}, {Farrens}, {Ferriol}, {Frailis},
  {Franceschi}, {Franzetti}, {Fumana}, {Garilli}, {Gillard}, {Gillis},
  {Giocoli}, {Grazian}, {Grupp}, {Guzzo}, {Haugan}, {Holmes}, {Hormuth},
  {Hornstrup}, {Jahnke}, {K{\"u}mmel}, {Kermiche}, {Kiessling}, {Kilbinger},
  {Kohley}, {Kunz}, {Kurki-Suonio}, {Ligori}, {Lilje}, {Lloro}, {Maiorano},
  {Mansutti}, {Marggraf}, {Markovic}, {Marulli}, {Massey}, {Maurogordato},
  {Medinaceli}, {Meneghetti}, {Merlin}, {Meylan}, {Moresco}, {Moscardini},
  {Munari}, {Niemi}, {Padilla}, {Paltani}, {Pasian}, {Pedersen}, {Pettorino},
  {Polenta}, {Poncet}, {Popa}, {Raison}, {Renzi}, {Rhodes}, {Riccio}, {Rix},
  {Romelli}, {Roncarelli}, {Rosset}, {Rossetti}, {Saglia}, {Sapone},
  {Sartoris}, {Schneider}, {Scodeggio}, {Secroun}, {Seidel}, {Sirignano},
  {Sirri}, {Stanco}, {Tallada-Cresp{\'\i}}, {Tavagnacco}, {Taylor}, {Tereno},
  {Toledo-Moreo}, {Torradeflot}, {Tutusaus}, {Valentijn}, {Valenziano},
  {Vassallo}, {Wang}, {Zacchei}, {Zamorani}, {Zoubian}, {Andreon}, {Boucaud},
  {Colodro-Conde}, {Di Ferdinando}, {Graci{\'a}-Carpio}, {Lindholm}, {Maino},
  {Mei}, {Scottez}, {Sureau}, {Tenti}, {Zucca}, {Borlaff}, {Ballardini},
  {Biviano}, {Bozzo}, {Burigana}, {Cabanac}, {Cappi}, {Carvalho}, {Casas},
  {Castignani}, {Cooray}, {Coupon}, {Courtois}, {Cuby}, {Davini}, {De Lucia},
  {Desprez}, {Dole}, {Escartin}, {Escoffier}, {Farina}, {Fotopoulou}, {Ganga},
  {Garcia-Bellido}, {George}, {Giacomini}, {Gozaliasl}, {Hildebrandt}, {Hook},
  {Huertas-Company}, {Kansal}, {Keihanen}, {Kirkpatrick}, {Loureiro},
  {Mac{\'\i}as-P{\'e}rez}, {Magliocchetti}, {Mainetti}, {Marcin}, {Martinelli},
  {Martinet}, {Metcalf}, {Monaco}, {Morgante}, {Nadathur}, {Nucita},
  {Patrizii}, {Peel}, {Potter}, {Pourtsidou}, {P{\"o}ntinen}, {Reimberg},
  {S{\'a}nchez}, {Sakr}, {Schirmer}, {Sefusatti}, {Sereno}, {Stadel},
  {Teyssier}, {Valieri}, {Valiviita}, \&
  {Viel}}]{EuclidCollaboration2022arXiv220614944E}
{Euclid Collaboration}, {Bisigello}, L., {Conselice}, C.~J., {et~al.} 2022,
  arXiv e-prints, arXiv:2206.14944

\bibitem[{{Gal} \& {Ghahramani}(2015)}]{2015arXiv150602142G}
{Gal}, Y. \& {Ghahramani}, Z. 2015, arXiv e-prints, arXiv:1506.02142

\bibitem[{{Ghirardini} {et~al.}(2022){Ghirardini}, {Bahar}, {Bulbul}, {Liu},
  {Clerc}, {Pacaud}, {Comparat}, {Liu}, {Ramos-Ceja}, {Hoang}, {Ider-Chitham},
  {Klein}, {Merloni}, {Nandra}, {Ota}, {Predehl}, {Reiprich}, {Sanders}, \&
  {Schrabback}}]{Ghirardini2022}
{Ghirardini}, V., {Bahar}, Y.~E., {Bulbul}, E., {et~al.} 2022, \aap, 661, A12

\bibitem[{{Grandis} {et~al.}(2021){Grandis}, {Bocquet}, {Mohr}, {Klein}, \&
  {Dolag}}]{2021MNRAS.507.5671G}
{Grandis}, S., {Bocquet}, S., {Mohr}, J.~J., {Klein}, M., \& {Dolag}, K. 2021,
  \mnras, 507, 5671

\bibitem[{Grandis {et~al.}(2019)Grandis, Mohr, Dietrich, Bocquet, Saro, Klein,
  Paulus, \& Capasso}]{Grandis:2018mle}
Grandis, S., Mohr, J.~J., Dietrich, J.~P., {et~al.} 2019, Mon. Not. Roy.
  Astron. Soc., 488, 2041

\bibitem[{{Green} {et~al.}(2019){Green}, {Ntampaka}, {Nagai}, {Lovisari},
  {Dolag}, {Eckert}, \& {ZuHone}}]{Green_2019}
{Green}, S.~B., {Ntampaka}, M., {Nagai}, D., {et~al.} 2019, \apj, 884, 33

\bibitem[{Harris {et~al.}(2020)Harris, Millman, van~der Walt, Gommers,
  Virtanen, Cournapeau, Wieser, Taylor, Berg, Smith, Kern, Picus, Hoyer, van
  Kerkwijk, Brett, Haldane, del R{\'{i}}o, Wiebe, Peterson,
  G{\'{e}}rard-Marchant, Sheppard, Reddy, Weckesser, Abbasi, Gohlke, \&
  Oliphant}]{harris2020array}
Harris, C.~R., Millman, K.~J., van~der Walt, S.~J., {et~al.} 2020, Nature, 585,
  357

\bibitem[{Ho {et~al.}(2021)Ho, Farahi, Rau, \& Trac}]{Ho_2021}
Ho, M., Farahi, A., Rau, M.~M., \& Trac, H. 2021, The Astrophysical Journal,
  908, 204

\bibitem[{Ho {et~al.}(2022)Ho, Ntampaka, Rau, Chen, Lansberry, Ruehle, \&
  Trac}]{Ho_2022}
Ho, M., Ntampaka, M., Rau, M.~M., {et~al.} 2022, Nature Astronomy, 6, 936

\bibitem[{{Ho} {et~al.}(2019){Ho}, {Rau}, {Ntampaka}, {Farahi}, {Trac}, \&
  {P{\'o}czos}}]{Ho_2019}
{Ho}, M., {Rau}, M.~M., {Ntampaka}, M., {et~al.} 2019, \apj, 887, 25

\bibitem[{{Ho} {et~al.}(2023){Ho}, {Soltis}, {Farahi}, {Nagai}, {Evrard}, \&
  {Ntampaka}}]{2023arXiv230300005H}
{Ho}, M., {Soltis}, J., {Farahi}, A., {et~al.} 2023, arXiv e-prints,
  arXiv:2303.00005

\bibitem[{Hunter(2007)}]{Hunter:2007}
Hunter, J.~D. 2007, Computing in Science \& Engineering, 9, 90

\bibitem[{{Klein} {et~al.}(2022){Klein}, {Oguri}, {Mohr}, {Grandis},
  {Ghirardini}, {Liu}, {Liu}, {Bulbul}, {Wolf}, {Comparat}, {Ramos-Ceja},
  {Buchner}, {Chiu}, {Clerc}, {Merloni}, {Miyatake}, {Miyazaki}, {Okabe},
  {Ota}, {Pacaud}, {Salvato}, \& {Driver}}]{Klein2022}
{Klein}, M., {Oguri}, M., {Mohr}, J.~J., {et~al.} 2022, \aap, 661, A4

\bibitem[{{Kodi Ramanah} {et~al.}(2020){Kodi Ramanah}, {Wojtak}, {Ansari},
  {Gall}, \& {Hjorth}}]{Kodi_Ramanah_2020}
{Kodi Ramanah}, D., {Wojtak}, R., {Ansari}, Z., {Gall}, C., \& {Hjorth}, J.
  2020, \mnras, 499, 1985

\bibitem[{Krizhevsky {et~al.}(2017)Krizhevsky, Sutskever, \&
  Hinton}]{krizhevsky2017imagenet}
Krizhevsky, A., Sutskever, I., \& Hinton, G.~E. 2017, Communications of the
  ACM, 60, 84

\bibitem[{{Lakshminarayanan} {et~al.}(2016){Lakshminarayanan}, {Pritzel}, \&
  {Blundell}}]{lakshminarayanan1612simple}
{Lakshminarayanan}, B., {Pritzel}, A., \& {Blundell}, C. 2016, arXiv e-prints,
  arXiv:1612.01474

\bibitem[{{Liu} {et~al.}(2022{\natexlab{a}}){Liu}, {Bulbul}, {Ghirardini},
  {Liu}, {Klein}, {Clerc}, {{\"O}zsoy}, {Ramos-Ceja}, {Pacaud}, {Comparat},
  {Okabe}, {Bahar}, {Biffi}, {Brunner}, {Br{\"u}ggen}, {Buchner}, {Ider
  Chitham}, {Chiu}, {Dolag}, {Gatuzz}, {Gonzalez}, {Hoang}, {Lamer}, {Merloni},
  {Nandra}, {Oguri}, {Ota}, {Predehl}, {Reiprich}, {Salvato}, {Schrabback},
  {Sanders}, {Seppi}, \& {Thibaud}}]{liu-efeds-clusters}
{Liu}, A., {Bulbul}, E., {Ghirardini}, V., {et~al.} 2022{\natexlab{a}}, \aap,
  661, A2

\bibitem[{{Liu} {et~al.}(2022{\natexlab{b}}){Liu}, {Merloni}, {Comparat},
  {Nandra}, {Sanders}, {Lamer}, {Buchner}, {Dwelly}, {Freyberg}, {Malyali},
  {Georgakakis}, {Salvato}, {Brunner}, {Brusa}, {Klein}, {Ghirardini}, {Clerc},
  {Pacaud}, {Bulbul}, {Liu}, {Schwope}, {Robrade}, {Wilms}, {Dauser},
  {Ramos-Ceja}, {Reiprich}, {Boller}, \& {Wolf}}]{Liu:2021idr}
{Liu}, T., {Merloni}, A., {Comparat}, J., {et~al.} 2022{\natexlab{b}}, \aap,
  661, A27

\bibitem[{{Mamon} {et~al.}(2013){Mamon}, {Biviano}, \&
  {Bou{\'e}}}]{2013MNRAS.429.3079M}
{Mamon}, G.~A., {Biviano}, A., \& {Bou{\'e}}, G. 2013, \mnras, 429, 3079

\bibitem[{{Mantz} {et~al.}(2015){Mantz}, {von der Linden}, {Allen},
  {Applegate}, {Kelly}, {Morris}, {Rapetti}, {Schmidt}, {Adhikari}, {Allen},
  {Burchat}, {Burke}, {Cataneo}, {Donovan}, {Ebeling}, {Shandera}, \&
  {Wright}}]{Mantz2015}
{Mantz}, A.~B., {von der Linden}, A., {Allen}, S.~W., {et~al.} 2015, \mnras,
  446, 2205

\bibitem[{{Merloni} {et~al.}(2012){Merloni}, {Predehl}, {Becker},
  {B{\"o}hringer}, {Boller}, {Brunner}, {Brusa}, {Dennerl}, {Freyberg},
  {Friedrich}, {Georgakakis}, {Haberl}, {Hasinger}, {Meidinger}, {Mohr},
  {Nandra}, {Rau}, {Reiprich}, {Robrade}, {Salvato}, {Santangelo}, {Sasaki},
  {Schwope}, {Wilms}, \& {German eROSITA Consortium}}]{2012arXiv1209.3114M}
{Merloni}, A., {Predehl}, P., {Becker}, W., {et~al.} 2012, arXiv e-prints,
  arXiv:1209.3114

\bibitem[{Ntampaka {et~al.}(2015)Ntampaka, Trac, Sutherland, Battaglia, P{\'{o}
  }czos, \& Schneider}]{Ntampaka_2015}
Ntampaka, M., Trac, H., Sutherland, D.~J., {et~al.} 2015, The Astrophysical
  Journal, 803, 50

\bibitem[{{Ntampaka} {et~al.}(2019){Ntampaka}, {ZuHone}, {Eisenstein}, {Nagai},
  {Vikhlinin}, {Hernquist}, {Marinacci}, {Nelson}, {Pakmor}, {Pillepich},
  {Torrey}, \& {Vogelsberger}}]{ntampaka1}
{Ntampaka}, M., {ZuHone}, J., {Eisenstein}, D., {et~al.} 2019, \apj, 876, 82

\bibitem[{{Old} {et~al.}(2014){Old}, {Skibba}, {Pearce}, {Croton}, {Muldrew},
  {Mu{\~n}oz-Cuartas}, {Gifford}, {Gray}, {von der Linden}, {Mamon},
  {Merrifield}, {M{\"u}ller}, {Pearson}, {Ponman}, {Saro}, {Sepp}, {Sif{\'o}n},
  {Tempel}, {Tundo}, {Wang}, \& {Wojtak}}]{2014MNRAS.441.1513O}
{Old}, L., {Skibba}, R.~A., {Pearce}, F.~R., {et~al.} 2014, \mnras, 441, 1513

\bibitem[{{Old} {et~al.}(2015){Old}, {Wojtak}, {Mamon}, {Skibba}, {Pearce},
  {Croton}, {Bamford}, {Behroozi}, {de Carvalho}, {Mu{\~n}oz-Cuartas},
  {Gifford}, {Gray}, {von der Linden}, {Merrifield}, {Muldrew}, {M{\"u}ller},
  {Pearson}, {Ponman}, {Rozo}, {Rykoff}, {Saro}, {Sepp}, {Sif{\'o}n}, \&
  {Tempel}}]{2015MNRAS.449.1897O}
{Old}, L., {Wojtak}, R., {Mamon}, G.~A., {et~al.} 2015, \mnras, 449, 1897

\bibitem[{Pedregosa {et~al.}(2011)Pedregosa, Varoquaux, Gramfort, Michel,
  Thirion, Grisel, Blondel, Prettenhofer, Weiss, Dubourg, Vanderplas, Passos,
  Cournapeau, Brucher, Perrot, \& Duchesnay}]{scikit-learn}
Pedregosa, F., Varoquaux, G., Gramfort, A., {et~al.} 2011, Journal of Machine
  Learning Research, 12, 2825

\bibitem[{{Planck Collaboration} {et~al.}(2020){Planck Collaboration},
  {Aghanim}, {Akrami}, {Ashdown}, {Aumont}, {Baccigalupi}, {Ballardini},
  {Banday}, {Barreiro}, {Bartolo}, {Basak}, {Battye}, {Benabed}, {Bernard},
  {Bersanelli}, {Bielewicz}, {Bock}, {Bond}, {Borrill}, {Bouchet}, {Boulanger},
  {Bucher}, {Burigana}, {Butler}, {Calabrese}, {Cardoso}, {Carron},
  {Challinor}, {Chiang}, {Chluba}, {Colombo}, {Combet}, {Contreras}, {Crill},
  {Cuttaia}, {de Bernardis}, {de Zotti}, {Delabrouille}, {Delouis}, {Di
  Valentino}, {Diego}, {Dor{\'e}}, {Douspis}, {Ducout}, {Dupac}, {Dusini},
  {Efstathiou}, {Elsner}, {En{\ss}lin}, {Eriksen}, {Fantaye}, {Farhang},
  {Fergusson}, {Fernandez-Cobos}, {Finelli}, {Forastieri}, {Frailis},
  {Fraisse}, {Franceschi}, {Frolov}, {Galeotta}, {Galli}, {Ganga},
  {G{\'e}nova-Santos}, {Gerbino}, {Ghosh}, {Gonz{\'a}lez-Nuevo}, {G{\'o}rski},
  {Gratton}, {Gruppuso}, {Gudmundsson}, {Hamann}, {Handley}, {Hansen},
  {Herranz}, {Hildebrandt}, {Hivon}, {Huang}, {Jaffe}, {Jones}, {Karakci},
  {Keih{\"a}nen}, {Keskitalo}, {Kiiveri}, {Kim}, {Kisner}, {Knox},
  {Krachmalnicoff}, {Kunz}, {Kurki-Suonio}, {Lagache}, {Lamarre}, {Lasenby},
  {Lattanzi}, {Lawrence}, {Le Jeune}, {Lemos}, {Lesgourgues}, {Levrier},
  {Lewis}, {Liguori}, {Lilje}, {Lilley}, {Lindholm}, {L{\'o}pez-Caniego},
  {Lubin}, {Ma}, {Mac{\'\i}as-P{\'e}rez}, {Maggio}, {Maino}, {Mandolesi},
  {Mangilli}, {Marcos-Caballero}, {Maris}, {Martin}, {Martinelli},
  {Mart{\'\i}nez-Gonz{\'a}lez}, {Matarrese}, {Mauri}, {McEwen}, {Meinhold},
  {Melchiorri}, {Mennella}, {Migliaccio}, {Millea}, {Mitra},
  {Miville-Desch{\^e}nes}, {Molinari}, {Montier}, {Morgante}, {Moss}, {Natoli},
  {N{\o}rgaard-Nielsen}, {Pagano}, {Paoletti}, {Partridge}, {Patanchon},
  {Peiris}, {Perrotta}, {Pettorino}, {Piacentini}, {Polastri}, {Polenta},
  {Puget}, {Rachen}, {Reinecke}, {Remazeilles}, {Renzi}, {Rocha}, {Rosset},
  {Roudier}, {Rubi{\~n}o-Mart{\'\i}n}, {Ruiz-Granados}, {Salvati}, {Sandri},
  {Savelainen}, {Scott}, {Shellard}, {Sirignano}, {Sirri}, {Spencer},
  {Sunyaev}, {Suur-Uski}, {Tauber}, {Tavagnacco}, {Tenti}, {Toffolatti},
  {Tomasi}, {Trombetti}, {Valenziano}, {Valiviita}, {Van Tent}, {Vibert},
  {Vielva}, {Villa}, {Vittorio}, {Wandelt}, {Wehus}, {White}, {White},
  {Zacchei}, \& {Zonca}}]{2020A&A...641A...6P}
{Planck Collaboration}, {Aghanim}, N., {Akrami}, Y., {et~al.} 2020, \aap, 641,
  A6

\bibitem[{{Predehl} {et~al.}(2021){Predehl}, {Andritschke}, {Arefiev},
  {Babyshkin}, {Batanov}, {Becker}, {B{\"o}hringer}, {Bogomolov}, {Boller},
  {Borm}, {Bornemann}, {Br{\"a}uninger}, {Br{\"u}ggen}, {Brunner}, {Brusa},
  {Bulbul}, {Buntov}, {Burwitz}, {Burkert}, {Clerc}, {Churazov}, {Coutinho},
  {Dauser}, {Dennerl}, {Doroshenko}, {Eder}, {Emberger}, {Eraerds},
  {Finoguenov}, {Freyberg}, {Friedrich}, {Friedrich}, {F{\"u}rmetz},
  {Georgakakis}, {Gilfanov}, {Granato}, {Grossberger}, {Gueguen}, {Gureev},
  {Haberl}, {H{\"a}lker}, {Hartner}, {Hasinger}, {Huber}, {Ji}, {Kienlin},
  {Kink}, {Korotkov}, {Kreykenbohm}, {Lamer}, {Lomakin}, {Lapshov}, {Liu},
  {Maitra}, {Meidinger}, {Menz}, {Merloni}, {Mernik}, {Mican}, {Mohr},
  {M{\"u}ller}, {Nandra}, {Nazarov}, {Pacaud}, {Pavlinsky}, {Perinati},
  {Pfeffermann}, {Pietschner}, {Ramos-Ceja}, {Rau}, {Reiffers}, {Reiprich},
  {Robrade}, {Salvato}, {Sanders}, {Santangelo}, {Sasaki}, {Scheuerle},
  {Schmid}, {Schmitt}, {Schwope}, {Shirshakov}, {Steinmetz}, {Stewart},
  {Str{\"u}der}, {Sunyaev}, {Tenzer}, {Tiedemann}, {Tr{\"u}mper}, {Voron},
  {Weber}, {Wilms}, \& {Yaroshenko}}]{2021A&A...647A...1P}
{Predehl}, P., {Andritschke}, R., {Arefiev}, V., {et~al.} 2021, \aap, 647, A1

\bibitem[{{Ramos-Ceja} {et~al.}(2022){Ramos-Ceja}, {Oguri}, {Miyazaki},
  {Ghirardini}, {Chiu}, {Okabe}, {Liu}, {Schrabback}, {Akino}, {Bahar},
  {Bulbul}, {Clerc}, {Comparat}, {Grandis}, {Klein}, {Lin}, {Merloni},
  {Mitsuishi}, {Miyatake}, {More}, {Nandra}, {Nishizawa}, {Ota}, {Pacaud},
  {Reiprich}, \& {Sanders}}]{Ramos-Ceja2022}
{Ramos-Ceja}, M.~E., {Oguri}, M., {Miyazaki}, S., {et~al.} 2022, \aap, 661, A14

\bibitem[{{Scheck} {et~al.}(2023){Scheck}, {Sanders}, {Biffi}, {Dolag},
  {Bulbul}, \& {Liu}}]{Scheck2023}
{Scheck}, D., {Sanders}, J.~S., {Biffi}, V., {et~al.} 2023, \aap, 670, A33

\bibitem[{{Seppi} {et~al.}(2022){Seppi}, {Comparat}, {Bulbul}, {Nandra},
  {Merloni}, {Clerc}, {Liu}, {Ghirardini}, {Liu}, {Salvato}, {Sanders},
  {Wilms}, {Dwelly}, {Dauser}, {K{\"o}nig}, {Ramos-Ceja}, {Garrel}, \&
  {Reiprich}}]{Seppi2022}
{Seppi}, R., {Comparat}, J., {Bulbul}, E., {et~al.} 2022, \aap, 665, A78

\bibitem[{{Sunyaev} {et~al.}(2021){Sunyaev}, {Arefiev}, {Babyshkin},
  {Bogomolov}, {Borisov}, {Buntov}, {Brunner}, {Burenin}, {Churazov},
  {Coutinho}, {Eder}, {Eismont}, {Freyberg}, {Gilfanov}, {Gureyev}, {Hasinger},
  {Khabibullin}, {Kolmykov}, {Komovkin}, {Krivonos}, {Lapshov}, {Levin},
  {Lomakin}, {Lutovinov}, {Medvedev}, {Merloni}, {Mernik}, {Mikhailov},
  {Molodtsov}, {Mzhelsky}, {M{\"u}ller}, {Nandra}, {Nazarov}, {Pavlinsky},
  {Poghodin}, {Predehl}, {Robrade}, {Sazonov}, {Scheuerle}, {Shirshakov},
  {Tkachenko}, \& {Voron}}]{Sunyaev2021}
{Sunyaev}, R., {Arefiev}, V., {Babyshkin}, V., {et~al.} 2021, \aap, 656, A132

\bibitem[{Virtanen {et~al.}(2020)Virtanen, Gommers, Oliphant, Haberland, Reddy,
  Cournapeau, Burovski, Peterson, Weckesser, Bright, {van der Walt}, Brett,
  Wilson, Millman, Mayorov, Nelson, Jones, Kern, Larson, Carey, Polat, Feng,
  Moore, {VanderPlas}, Laxalde, Perktold, Cimrman, Henriksen, Quintero, Harris,
  Archibald, Ribeiro, Pedregosa, {van Mulbregt}, \& {SciPy 1.0
  Contributors}}]{2020SciPy-NMeth}
Virtanen, P., Gommers, R., Oliphant, T.~E., {et~al.} 2020, Nature Methods, 17,
  261

\bibitem[{Wadekar {et~al.}(2022{\natexlab{a}})Wadekar, Thiele, Hill, Pandey,
  Villaescusa-Navarro, Spergel, Cranmer, Nagai, Anglés-Alcázar, Ho, \&
  Hernquist}]{2209.02075}
Wadekar, D., Thiele, L., Hill, J.~C., {et~al.} 2022{\natexlab{a}}, The SZ
  flux-mass ($Y$-$M$) relation at low halo masses: improvements with symbolic
  regression and strong constraints on baryonic feedback

\bibitem[{Wadekar {et~al.}(2022{\natexlab{b}})Wadekar, Thiele,
  Villaescusa-Navarro, Hill, Cranmer, Spergel, Battaglia, Anglés-Alcázar,
  Hernquist, \& Ho}]{2201.01305}
Wadekar, D., Thiele, L., Villaescusa-Navarro, F., {et~al.} 2022{\natexlab{b}},
  Augmenting astrophysical scaling relations with machine learning :
  application to reducing the SZ flux-mass scatter

\bibitem[{{W}es {M}c{K}inney(2010)}]{mckinney-proc-scipy-2010}
{W}es {M}c{K}inney. 2010, in {P}roceedings of the 9th {P}ython in {S}cience
  {C}onference, ed. {S}t\'efan van~der {W}alt \& {J}arrod {M}illman, 56 -- 61

\bibitem[{{Yan} {et~al.}(2020){Yan}, {Mead}, {Van Waerbeke}, {Hinshaw}, \&
  {McCarthy}}]{Yan_2020}
{Yan}, Z., {Mead}, A.~J., {Van Waerbeke}, L., {Hinshaw}, G., \& {McCarthy},
  I.~G. 2020, \mnras, 499, 3445

\bibitem[{{ZuHone} {et~al.}(2022){ZuHone}, {Bahar}, {Biffi}, {Dolag},
  {Sanders}, {Bulbul}, {Liu}, {Dauser}, {K{\"o}nig}, {Zhang}, \&
  {Ghirardini}}]{ZuHone2022}
{ZuHone}, J., {Bahar}, Y.~E., {Biffi}, V., {et~al.} 2022, arXiv e-prints,
  arXiv:2212.11028

\end{thebibliography}

\appendix
\section{Gaussian Kernel}
\label{app:data}
X-ray astronomical maps correspond to the spatial distribution of the detected photons. Depending on the source luminosity, the amount of detected photons emitted by the source can be very sparse and can lead to a difficult recognition of the source shape. In fact, some EBI contain only a couple of source photons. In the EBI generation, we use Gaussian blur to support the learning process of our models. Gaussian blur is usually used in image processing to reduce the noise and detail. In our case, our aim is to smooth the sparse distribution of detected photons.

We utilize the kernel operation:
\begin{equation}
    \tilde{I}(x,y,E) = \sum_{i=-X}^{X}\sum_{j=-Y}^{Y}\sum_{k=-Z}^{Z}I(x-i,y-j,E-k)G(i,j,k)~,
    \label{eq:kernelsmoothing}
\end{equation}
 where $I$ denotes our original EBI-array and $\tilde{I}$ the smoothed array. For the convolution, we use the following kernel $G(i,j,k)$
\begin{equation}
G(i,j,k)=\frac{1}{N}\exp\left( -\frac{(i/X)^2+(j/Y)^2+(k/Z)^2}{2\sigma^2} \right)\,,
\label{eq:gaussian}
\end{equation}
where $N$ is the normalizing factor and $\sigma$ the distributions standard deviation.  
We observed that smoothing also along the channel dimension, enhances the source structure in each channel of our EBIs and leads to improvements in our models' performance. The final choice of the filter size is $(3,3,11)$ which corresponds to $(X,Y,Z)=(1, 1, 5)$ in~\eqref{eq:kernelsmoothing} and a standard deviation of $\sigma=0.75$. We use \texttt{scipy.signal.convolve} \citep{2020SciPy-NMeth} to perform this smoothing.

\section{Hyperparameters}
\label{app:hyperparameter}
To identify well-performing neural networks we have performed a hyper-parameter search on which we provide an overview in this appendix. 
\begin{itemize}
    \item {\bf CNN hyperparameters:} We run standard variations of the kernel size, and number of filters in convolutional blocks. We found that average pooling worked better than maxpooling. Further we varied the activation functions (relu and leaky relu), and the number and dimensions of the final dense layers.
    \item {\bf Different network architectures:} We have also observed that locally connected convolutional layers did not increase the performance. They were comparable to our CNN approach. In our hyper-parameter analysis we have compare the behaviour of different pooling layers, and have varied the size and number of hidden layers, kernel sizes moderately. We find for a range of hyperparameters good performance.
   
    \item {\bf EBI hyperparameters:} We have varied the extraction size among $100,$ $200,$ $300,$ and $500$ pixels. $300$ pixels showed the best performance and it corresponds to a size where essentially all cluster photons are contained within the extraction range. We also experimented with larger extraction sizes and could not find an improvement in performance. We found no difference between using the eSASS cluster center or the cluster center provided from the simulations as the center for our EBI image, i.e.~training with either of them resulted in no difference in performance. We also tried rescaling the EBI images according to redshift but could not identify better performance.
    \item {\bf Regression losses:} Besides our likelihood loss we have experimented with various standard regression losses (mean squared error, mean average percentage error, and mean squared logarithmic error). We find that they are generally lead to similar behaviour on the mass scatter.
    \item {\bf Classification vs.~regression}: As classification tasks are sometimes easier learning problems than regression in machine learning, we have experimented with classification where the classes correspond to different mass bins. It turned out that this approach did not lead to improved performance in comparison to our current CNN-based approach.
    \item {\bf Optimizers:} We have used Adam and generally found reducing the learning rate on plateau to be useful for performance and scanned through minimal learning rates. A batch size of a $100$ was chosen, with variations not showing an increased performance.
\end{itemize}
Overall we identify in this search several models with different hyperparameters which do perform similarly. We think that this robustness is encouraging for these NN approaches.

\end{document}